\documentclass[12pt,a4paper]{article}
\usepackage{cite}
\usepackage{mathtext}
\usepackage{hyperref}
\usepackage{amssymb,amsthm,amsmath}
\usepackage{array}
\usepackage[title,titletoc]{appendix}

\begin{document}
\author{\bf Yu.A. Markov\thanks{e-mail:markov@icc.ru},\,
M.A. Markova\thanks{e-mail:markova@icc.ru}, and A.I. Bondarenko\thanks{e-mail:370omega@mail.ru}}
\title{Path integral representation\\ for inverse third order wave operator\\
within the Duffin-Kemmer-Petiau formalism. \!\!I}
%
%
\date{\it\normalsize
\begin{itemize}
\item[]
Matrosov Institute for System Dynamics and Control Theory SB RAS
Irkutsk, Russia
\end{itemize}}
\thispagestyle{empty}
\maketitle{}


\def\theequation{\arabic{section}.\arabic{equation}}
\vspace{-0.2cm}
\[
{\bf Abstract}
\]
{Within the framework of the Duffin-Kemmer-Petiau (DKP) formalism with a deformation, an approach to the construction of the path integral representation in parasuperspace for the Green's function of a spin-1 massive particle in external Maxwell's field is developed. For this purpose a connection between the deformed DKP-algebra and an extended system of the parafermion trilinear commutation relations for the creation and annihilation operators $a^{\pm}_{k}$ and for an additional operator $a_{0}$ obeying para-Fermi statistics of order 2 based on the Lie algebra $\mathfrak{so}(2M+2)$ is established. The representation for the operator $a_{0}$ in terms of generators of the orthogonal group $SO(2M)$ correctly reproducing action of this operator on the state vectors of Fock space is obtained. An appropriate system of the parafermion coherent states as functions of para-Grassmann numbers is introduced. The procedure of the construction of finite-multiplicity approximation for determination of the path integral in the relevant phase space is defined through insertion in the kernel of the evolution operator with respect to para-supertime of resolutions of the identity. In the basis of parafermion coherent states a matrix element of the contribution linear in covariant derivative $\hat{D}_{\mu}$ to the time-dependent Hamilton operator $\hat{\cal H}(\tau)$, is calculated in an explicit form. For this purpose the matrix elements of the operators $a^{\phantom{2}}_0$, $a_{0}^{2}$, the commutators $[\hspace{0.03cm}a^{\phantom{\pm}\!}_{0}, a^{\pm}_{n}\hspace{0.02cm}]$, $[\hspace{0.03cm}a^{2}_{0}, a^{\pm}_{n}\hspace{0.02cm}]$, and the product $\hat{A}\hspace{0.03cm}[\hspace{0.03cm}a^{\phantom{\pm}\!}_{0}, a^{\pm}_{n}\hspace{0.02cm}]$ with $\hat{A} \equiv\exp\hspace{0.02cm}\bigl(-i\frac{2\pi}{3}\,a_{0}\bigr)$, were preliminary defined.
}
{}


\newpage

\section{Introduction}
\setcounter{equation}{0}
\label{section_1}

The propagators (the Green's functions) for free quantized fields involved in the interaction processes and their generalization to the case of external classical fields in a system are important structural elements in the calculation of the Feynman diagrams in quantum field theory. However, in a number of problems it is convenient to have an alternative to the standard technique in quantum field theory. One of such alternatives is a possibility to present the Green's functions in the form of quantum-mechanical path integrals and thereby to reformulate quantum field theory in the language of world-lines of particles.\\
\indent The representations in the form of path integrals were constructed for the scalar propagator \cite{feynman_1950}, the electron propagator in an external Maxwell field \cite{fradkin_1966, batalin_1970, bordi_1980, henty_1988, cooper_1988, fainberg_1988, fradkin(1)_1991, gitman_1997} and for the quark propagator in an external Yang-Mills gauge field \cite{halpern_1977, borisov_1982, shvartsman_1990, balitsky_2001}. In constructing the desired representations the variety of approaches and methods was used. The case of propagators for particles with half-integer spin (electrons and quarks) in external gauge fields and also their generalization to the case of supersymmetric theories \cite{duncan_1986, fradkin(2)_1991, grundberg_1993, marnelius_1994(1)} were studied in greater detail. We note that the representations of the Green's functions  (and the one-loop effective actions closely connected with them) in the form of path integrals enable one to obtain by a more simple way some well-known results of quantum field theory and in particular, of quantum electrodynamics, for example, the Euler-Heisenberg Lagrangian for the case of strong constant or slowly varying  field \cite{schmidt_1993}. Moreover, this approach was successfully used in calculating the two-loop effective action that enables one to calculate a correction to the effective Euler-Heisenberg Lagrangian \cite{reuter_1997, sato_1998, sato_1999}. Finally, the exact calculation of functional integrals for special configurations of external fields gives an alternative possibility to study a problem of vacuum stability perturbed by external Maxwell's or Yang-Mills' fields.\\
\indent Whereas in principle, one can construct the representation in the form of path integral for propagators of free fields with an arbitrary spin, such an attempt for fields with a spin, which is greater than $1/2$ interacting with an external (Abelian or non-Abelian) gauge field encounters a problem of consistency \cite{vijayalakshmi_1979, howe_1989, buchbinder_1993, gitman_1995}. In future, we focus on the propagator of a field with the spin 1, more exactly, on the propagator of a charged massive vector particle in the external Maxwell's field.\\
\indent In this paper we would like to propose an approach to the construction of the representation for the Green's function of a vector particle in an external field in the form of path integral based on a well-known Duffin-Kemmer-Petiau (DKP) formalism \cite{duffin_1938, kemmer_1939, petiau_1936} developed for describing relativistic scalar and vector particles.  In the four-dimensional Euclidean space-time the ten-dimensional representation of the DKP-algebra corresponds to fields with spin 1. One of the most important advantages of this formalism is a possibility of using a well-developed technique for the case of the electron and quark propagators. In constructing such a representation for the vector particle we will follow mainly approaches suggested by Halpern, Jevicki and Senjanovi\'c \cite{halpern_1977}, Borisov and Kulish \cite{borisov_1982}, Fradkin and Shvartsman \cite{fradkin_1988}, Fradkin and Gitman \cite{fradkin(1)_1991} and van Holten \cite{holten_1995}. We study in more detail a connection between para-Fermi quantization based on the Lie algebra of the orthogonal group $SO(2M+2)$ and the Duffin-Kemmer-Petiau theory with a deformation early suggested in \cite{markov_2015}, where as the deformation parameter a primitive cubic root of unity is used and the wave function of the particle with spin 1 obeys the third order wave equation. Note that an analysis of this connection is of particular mathematical interest without an application to a specific physical problem, since the connection represents nontrivial synthesis of various subjects such as algebra, the theory of classical Lie groups and theoretical aspects of (para)quantization of fields. The analysis will be considered in details in our separate work \cite{markov_2020}. In the present paper we will mainly use only the resulting formulas 
from \cite{markov_2020}.\\
\indent Relativistic particle theories with spin 1 were studied since that time, when Dirac has written out his famous equation for a particle with spin $1/2$ \cite{pauli_1934, dirac_1936, proca_1936, kemmer_1938, kemmer_1939, bhabha_1938, schrodinger_1943(1), schrodinger_1943(2), schrodinger_1955}. In particular, it was shown that the well-known Proca equation for a massive vector field can be rewritten in the matrix form of DKP-relativistic wave equation. The description of spin degree of freedom of a massive non-Abelian vector field based on DKP-approach can be found in the papers by Bogush and Zhirkov \cite{bogush_1977}, Okubo and Tosa \cite{okubo_1979}, and Gribov \cite{gribov_1999}.\\
\indent For the first time, the interaction with an external gauge (electromagnetic) field within the framework of the DKP-formalism was considered in the pioneering paper by Kemmer \cite{kemmer_1939}. The interaction with the external field was introduced within the framework of the minimal coupling scheme that thereby actually provides gauge invariance of the DKP Lagrangian. Further, in a number of papers \cite{nowakowski_1998, lunardi_2000, fainberg_2000} a question of the interaction of a charged vector particle with electromagnetic field was analyzed in more depth. In particular, it was explained that the main difference of the DKP-equation from the Dirac equation is that it involves redundant components. Some interaction terms in the Hamilton form of the DKP equation do not have a physical meaning and will not affect the calculation of physical observables. Furthermore, Nowakowski  \cite{nowakowski_1998} pointed out that the DKP-equation of the second order obtained by Kemmer \cite{kemmer_1939} by analogy with the second order Dirac equation has a rather limited physical applicability, since (1) it is only one member of a class of second order equations which can be derived from the original DKP-equation in external electromagnetic field and (2) it has not a back-transformation, which would allow us to obtain solutions of the first order DKP-equation from solutions of the second order equation as it is in the Dirac theory. These results are true for an arbitrary representation of $\beta$-matrices (even not necessarily irreducible). All these principal issues arising in the problem of interacting DKP-field with an external Abelian one (and also with non-Abelian one) would have to take into account in solving the problem stated in the present paper.\\
\indent Further, the Duffin-Kemmer-Petiau algebra is closely related to an entirely different branch of theoretical physics, namely, the theory of parastatistics, more exactly, to the para-Fermi statistics of order $p = 2$. This nontrivial fact  was noted for the first time in the papers by Volkov \cite{volkov_1959}, Chernikov \cite{chernikov_1962} and independently by Ryan and Sudarshan \cite{ryan_1963}. This connection provided an opportunity to present the DKP-algebra within the framework of an operator formalism (see section \ref{section_3}) in the form of parafermion algebra of order $p = 2$ and to realize a spin space of vector particle as a Fock space for a system of para-Fermi operators \cite{bogolyubov_1989}.\\
\indent However, a preliminary analysis \cite{markov_2018} has shown that the use of parafermion algebra in the standard form is insufficient for solving the stated problem and here, a generalization of this algebra  would be required. As is well known, trilinear commutation relations for the para-Fermi statistics generate algebra which is isomorphic to the Lie algebra $\mathfrak{so}(2M + 1)$ \cite{kamefuchi_1962}. Geyer in the paper \cite{geyer_1968}
has suggested to extend this isomorphism to the Lie algebra $\mathfrak{so}(2M+2)$. The extension is of great value  for us, since in the corresponding algebra of para-Fermi operators an additional operator $a_{0}$ arises. This operator in the case of parastatistics of order 2 can be related to within a sign to the Schr\"odinger  ``pseudomatrix'' $\omega$ \cite{schrodinger_1943(1)} playing a key role in constructing the divisor for the first order DKP operator of a vector particle in an external gauge field \cite{markov_2015}. This divisor enables us in particular, to write an operator expression for the inverse propagator of the vector particle in the form of the Fock-Schwinger proper parasupertime representation.\\
\indent There are a few papers, where a question of the construction of path integral for a system of identical particles obeying parastatistics was considered (see, e.g. Polychronakos \cite{polychronakos_1996}, Chaichian and Demichev \cite{chaichian_2001}, Greenberg and Mishra \cite{greenberg_2004}). In this direction of researches the papers by Omote and Kamefuchi \cite{omote_1979} and Ohnuki and Kamefuchi \cite{ohnuki_1980} are of particular interest for us. For a generalization of the notion of path integral to the case of parafermion variables in these papers the first step was to suggest an generalization of the well-known Grassmann algebra to the so-called  {\it para-Grassmann}  algebra \cite{kalnay_1976}. This generalization is a  direct analogue of generalization of the Fermi operators to the case of the para-Fermi operators in parastatistics. The authors have introduced the definition of the para-Grassmann algebra of arbitrary order $p$, the notions of integration and differentiation in this algebra, change of variables in integrals, Fourier transformation and so on. They also have defined the notions of coherent states for the para-Fermi operators and written out the formula of resolution of the identity (the completeness relation). These parafermion coherent states and resolution of the identity are of fundamental  importance in a procedure of the construction of path integrals. The authors have constructed the path integral for the para-Fermi fields using para-Grassmann variables following the  definition of the path integral as the limit of a product of time evolution operators for small time intervals. In formulating the theory the authors actively used the so-called Green ansatz \cite{green_1953}. Note that the papers \cite{omote_1979, ohnuki_1980} are a direct generalization of the paper by Ohnuku and Kashiwa \cite{ohnuki_1978}, in which the construction of path integrals over Grassmann variables was presented, and are decisive in solving the problem stated in the present work. Essentially all the mathematical apparatus constructed by these authors will be actively used in the suggested research.\\
\indent It should be also noted that there exists another direction of the description of massive and massless spinning particles within the framework of the so-called pseudoclassical mechanics using odd (``spinning'') Grassmann or para-Grassmann variables in addition to usual even variables (coordinate and momentum). The results of these researches are also important for us, since the Lagrangians analyzed there (and correspondingly, the classical actions) of free particles or particles in an external field, massive or massless ones possessing symmetries of various kinds, need to appear in one form or another in the exponential in the path integral representation of propagators of these particles in quantum field theory, thus forming a connection between relativistic mechanics of classical spinning particles and the Green's functions in quantum field theory.\\
\indent In the paper by Gershun and Tkach \cite{gershun_1979} in particular it was shown that for the description of classical and quantum dynamics of a particle with spin 1 it is necessary to introduce two real Grassmann-valued vector variables $\psi_{\mu}^{k},\, k=1, 2$ (instead of one variable as in the case of spin $1/2$). Superspace formulation of the given approach with the so-called doubly supersymmetry can be found in \cite{kowalski-glikman_1988, sorokin_1989, casana_2009}. Further, in the paper by Barducci and Lussanna \cite{barducci_1983(1)} the pseudoclassical description of a massless particle with helicity $\pm 1$ in terms of complex conjugate pair of Grassmann 4-vectors $\psi_{\mu}$ and $\psi_{\mu}^{*}$ was presented. With the use of canonical quantization, one-photon wave function in the Lorentz gauge was obtained and based on quantization within the framework of path integration non-covariant transverse propagator for a free field was derived. The authors have also considered the case of describing massive photon within the framework of pseudoclassical mechanics \cite{barducci_1983(2)}. They have suggested a set of a first-class constraints, which after quantization reproduce the Proca equation for a massive vector field.\\
\indent In two subsequent papers Gershun and Tkach \cite{gershun_1984, gershun_1985} have analysed  more closely a case of vector particles. It was cleared up that for a massless particle the descriptions by using a set of two Grassmann variables $\psi_{\mu}^{k}$ and with the help of one para-Grassmann variable $\psi_{\mu}$ of order $p = 2$ (i.e. $(\psi_{\mu})^3 = 0)$ are fully equivalent, whereas the description of a massive particle with the spin 1 is possible only with the para-Grassmann variables $\psi_{\mu}$ and $\psi_{5}$. The Lagrangian, which describes the motion of the free massive particle with spin 1 in terms of the para-Grassmann variables, has the following form:
\begin{equation}
L=L_{0}+L_{m},
\label{eq:1q}
\end{equation}
where
\begin{align}
&L_{0} = \displaystyle\frac{1}{2\hspace{0.02cm}e}\,\dot{x}_{\mu}^{2}
-
\displaystyle\frac{i}{2}\,[\hspace{0.02cm}\psi_{\mu},\dot{\psi}_{\mu}\hspace{0.02cm}]
-
\displaystyle\frac{i}{2\hspace{0.02cm}e}\,[\hspace{0.02cm}\lambda,
\dot{x}_{\mu}\hspace{0.02cm}\psi_{\mu}\hspace{0.02cm}]
-
\displaystyle\frac{1}{8\hspace{0.02cm}e}\,
[\hspace{0.02cm}\lambda,\psi_{\mu}\hspace{0.02cm}]^{\hspace{0.02cm}2}
+
B\hspace{0.02cm}[\hspace{0.02cm}\psi_{\mu}, \psi_{\mu}\hspace{0.02cm}]^{\hspace{0.02cm}2}\hspace{0.02cm}V,
\notag\\[1ex]
&L_{m} = \displaystyle\frac{e}{2}\,m^2
+
\displaystyle\frac{i}{2}\,[\hspace{0.02cm}\psi_{5\,},\dot{\psi_{5}}\hspace{0.02cm}]
+
\displaystyle\frac{i}{2}\,m\hspace{0.02cm}[\hspace{0.02cm}\lambda,\psi_{5}\hspace{0.02cm}]
-
2B\hspace{0.02cm}[\hspace{0.02cm}\psi_{\mu}, \psi_{\mu}\hspace{0.02cm}]\hspace{0.01cm}[\hspace{0.02cm}\psi_{5}, \psi_{5}\hspace{0.02cm}]\hspace{0.02cm}V.
\notag
\end{align}
Here, $\mu = 1,2,3,4$, the dot denotes differentiation with respect to $\tau$, the fields $e(\tau), \lambda(\tau)$ and $V(\tau)$ are (one-dimensional) vierbein, gravitino and vector fields, respectively, and play the role of the Lagrange multipliers. The Lagrangian is invariant up to a total derivative under the coordinate transformation of the parameter $\tau$, the infinitesimal supersymmetry transformations with an arbitrary Grassmann-valued function $\alpha = \alpha(\tau)$ and local $O(2)$ internal transformations. A set of the classical para-Grassmann variables $(\psi_{\mu},\psi_5)$ obeys trilinear relation
\[
\psi_{\mu}\psi_{\nu}\psi_{\lambda} + \psi_{\lambda}\psi_{\nu}\psi_{\mu} = 0,
\]
which after quantization passes into the operator relation of the algebra of para-Fermi fields\footnote{\,We have redefined the para-Grassmann numbers and operators from \cite{gershun_1985} as follows:\! $\psi_{\mu}\rightarrow\sqrt{2}\hspace{0.03cm}\psi_{\mu},\hspace{0.03cm} \lambda\rightarrow\sqrt{2}\hspace{0.03cm}\lambda$ etc.}
\[
\hat{\psi}_{\mu} \hat{\psi}_{\nu} \hat{\psi}_{\lambda} + \hat{\psi}_{\lambda } \hat{\psi}_{\nu} \hat{\psi}_{\mu}
=
\hbar\,(\delta_{\mu \nu} \hat{\psi}_{\lambda} + \delta_{\lambda \nu} \hat{\psi}_{\mu}),
\]
where now $\mu,\nu,\lambda = 1,2,3,4,5$. The pseudoclassical Lagrangian (\ref{eq:1q}) has a direct relationship to our problem, and therefore is of greater interest for us.\\
\indent In the papers by Korchemsky \cite{korchemski_1991, korchemski_1992}, the Lagrangian (\ref{eq:1q}) in the case, when $B = 0$ was used for the first quantization of a relativistic spinning particle. The author has shown that in the massless case, i.e. for $L_{m} = 0$, after quantization the physical subspace of the parasupersymmetric particle whose spinning coordinates belong to the irreducible representations of the Duffin-Kemmer-Petiau algebra labelled by integer number is described by the strength tensors of antisymmetric gauge fields and topological gauge fields.\\
\indent Marnelius and M{\aa}rtensson \cite{marnelius_1989}, Lin and Ni \cite{lin_1990}, Rivelles and Sandoval \cite{rivelles_1991} and Marnelius \cite{marnelius_1994(2)} have considered the BRST-quantization (within the framework of the Batalin-Fradkin-Vilkovisky procedure) of a model of relativistic spinning particle with $N = 2$ extended local supersymmetry on the worldline, which after quantization describes a particle with spin 1. Further, Gitman, Gon\c{c}alves and Tyutin \cite{gitman_1995} have suggested a consistent procedure for canonical quantization of the pseudoclassical model of a spin 1 relativistic particle. They have shown that the quantum mechanics obtained after quantization for the massive case is equivalent to the Proca theory, and for the massless case, to the Maxwell theory. In this paper the case of the interaction with an electromagnetic field was also considered and it was shown that for an arbitrary external field the corresponding Lagrange equations become inconsistent. Only in the case of a constant external field (the authors in particular have considered an external constant magnetic field) one can obtain the consistent equations of motion.\\
\indent A possibility of introducing the interaction with external electromagnetic field in the model with $N = 2$ extended supersymmetry on the worldline was also considered in the paper by P. Howe {\it et al.} \cite{howe_1989}. The authors have shown also impossibility of the self-consistent description of interaction of the charged vector particle with the electromagnetic field. In addition, it could, however, be said that pseudoclassical models for a particle with spin 1 admit the interaction with an external gravitation field \cite{buchbinder_1993, bastianelli_2005(1), bastianelli_2005(2)}.\\
\indent By this means within the framework of standard approaches such as the pseudoclassical mechanics, the usual Duffin-Kemmer-Petiau theory, an approach based on the Bargmann-Wigner equations and so on it is impossible  in a consistent manner to introduce the interaction of the charged vector particle with external gauge fields. Our approach will allow one to get around this problem by the increasing complexity of  the first order differential operator acting on a wave function of the vector particle.\\
\indent The paper is organized as follows. In section \ref{section_2}, a brief review of our work \cite{markov_2015} devoted to deriving the third order wave equation within the framework of Duffin-Kemmer-Petiau theory with a deformation, is presented. In section \ref{section_3}, for constructing the path integral representation we give all necessary formulas of operator formalism: the trilinear relations to which the operators of creation and annihilation of parafermions obey, the basis of parafermion coherent states in the spin space $L$, the normalization and completeness relations for the coherent states and so on. The generalized Hamilton operator $\hat{\cal H} = \hat{\cal H}(\tau)$ explicitly depending on the evolution parameter $\tau$ and containing linear, quadratic and cubic terms in the covariant derivative $\hat{D}_{\mu}$ is introduced. On the basis of the Hamiltonian the proper-time evolution operator $\hat{U}(T, 0)$ used in constricting the scheme of finite multiplicity approximations is defined. 
In section \ref{section_4}, the form of the initial for further analysis matrix element of contribution to the generalized Hamilton operator linear in covariant derivative is written out. In section \ref{section_5} a representation for the operator $a_{0}$ in terms of the generators of the group $SO(4)$ correctly reproducing action on the state vectors is suggested. A connection of this operator with the pseudoclassical DKP-operator $\hat{\omega}$ is obtained. The matrix element of the $a_{0}$ in the basis of parafermion coherent states is derived.\\
\indent Section \ref{section_6} is devoted to calculation of the matrix element for the Geyer operator $a_{0}^{2}$, an analysis of its structure and derivation of its more compact and visual representation. Section \ref{section_7} is concerned with deriving the matrix elements of the commutators 
$[\hspace{0.03cm}a_{0}, a^{\pm}_{n}\hspace{0.02cm}]$, $[\hspace{0.03cm}a^{2}_{0}, a^{\pm}_{n}\hspace{0.02cm}]$, which arise within the framework of finite-multiplicity approximation in constructing the required path integral representation of the Green's function for a vector particle. In section \ref{section_8} a similar calculation of the matrix elements of the product $\hat{A}\hspace{0.03cm}[\hspace{0.03cm}a_{0}, a^{\pm}_{n}\hspace{0.02cm}]$, where $\hat{A} \equiv\exp\hspace{0.02cm}\bigl(-i\frac{2\pi}{3}\,a_{0}\bigr)$, is performed. More compact representations for these matrix elements are defined. On the basis of the obtained expressions for the matrix elements in this and previous sections a complete expression for the matrix element $\langle\hspace{0.02cm}(k)^{\prime}_{p}\hspace{0.02cm}|\hspace{0.02cm} [\hspace{0.03cm}\chi,\hat{\cal L}(z,\hat{D})\hspace{0.01cm}]\hspace{0.01cm}| \hspace{0.02cm}(k - 1)_{x}\hspace{0.02cm}\rangle$ from section \ref{section_4}, is given. In the concluding section \ref{section_9} the key points of our work are specified.\\
\indent In Appendix \ref{appendix_A} the basic relations of the Lie algebra of the orthogonal group $SO(2M+2)$ are given. Appendix \ref{appendix_B} is devoted to the formulation of the definition of a para-Grassmann algebra in a spirit of the paper by Omote and Kamefuchi \cite{omote_1979}. The trilinear relations between the para-Grassmann numbers $\xi_{k}$ and the creation and annihilation para-Fermi operators $a_{n}^{\pm}$ of parastatistics are written out. Two necessary formulas of differentiation with respect to para-Grassmann variables are given.


\section{Third-order wave operator}
\setcounter{equation}{0}
\label{section_2}

As already mentioned in Introduction in the paper by Nowakowski \cite{nowakowski_1998} devoted to the problem of electromagnetic coupling in the Duffin-Kemmer-Petiau theory, unusual circumstance relating to a second order DKP equation has been pointed out. It is connected with the fact that the second order Kemmer equation \cite{kemmer_1939} lacks a back-transformation which would allow one to obtain solutions of the first order DKP equation from solutions of the second order equation, as is the case in Dirac's theory. The reason of the latter is that the Klein-Gordon-Fock divisor \cite{umezawa_1956(2), takahashi_book} in the spin-1 case\footnote{\,Henceforth, we put $\hbar\!=\!c\!=\!1$, use Euclidean metric $\delta_{\mu \nu} = {\rm diag}\hspace{0.03cm}(1, 1, \hspace{0.03cm}\ldots\hspace{0.03cm},1)$, and adopt the usual summation convention only over repeated Greek indices $\mu,\,\nu,\,\lambda,\,\ldots$. For Latin indices $k,\,l,\,m,\,\ldots$ we will use the summation sign explicitly.}
\[
d(\partial) = \frac{1}{m}\,(\hspace{0.02cm}\Box - m^{2}\hspace{0.02cm})I  + i\hspace{0.02cm}\beta_{\mu}\hspace{0.02cm}\partial_{\mu}
+
\frac{1}{m}\,\beta_{\mu}\beta_{\nu}\hspace{0.02cm}\partial_{\mu}\partial_{\nu}
\]
ceases to be commuted with the original DKP operator
\[
L(\partial) \equiv i\hspace{0.02cm}\beta_{\mu}\partial_{\mu} + m\hspace{0.02cm} I,
\]
when we introduce the interaction with an external electromagnetic field within the framework of the minimal coupling scheme: $\partial_{\mu}\rightarrow D_{\mu}\equiv\partial_{\mu} + i\hspace{0.015cm}eA_{\mu}$, i.e.
\[
[\hspace{0.04cm}d(D), L(D)\hspace{0.03cm}] \neq 0.
\]
Here, $I$ is the unity matrix; $\Box\equiv\partial_{\mu}\partial_{\mu},\ \partial_{\mu}\equiv \partial/\partial x_{\mu}$, and the matrices $\beta_{\mu}$ obey the famous trilinear relation
\begin{equation}
\beta_{\mu}\beta_{\nu}\beta_{\lambda} + \beta_{\lambda}\beta_{\nu}\beta_{\mu} =
\delta_{\mu\nu}\beta_{\lambda} + \delta_{\lambda\nu}\beta_{\mu}.
\label{eq:2q}
\end{equation}
One of the negative consequences of this fact is impossibility to construct the Green function representation of (massive) vector particle in an external gauge field in the form of path integral in a certain (para)superspace remaining only within the framework of the original DKP-theory.\\
\indent Nowakowski has suggested a way how this problem may be circumvented. To achieve the commutativity of the divisor $d(D)$ and of the DKP operator $L(D)$ in the presence of an external electromagnetic field we have to give up the requirement that the product of these two operators is an operator of the Klein-Gordon-Fock type, i.e.
\[
d(D)L(D) \neq  (D^2 - m^2)I + {\cal G}\hspace{0.02cm}[A_{\mu}],
\]
where ${\cal G}\hspace{0.02cm}[A_{\mu}]$ is a functional of the potential $A_{\mu}$, which vanishes in the absence of interaction. In other words it is necessary to introduce into consideration not the second order, but a higher order wave equation which would have the same virtue as the second order Dirac equation, i.e. a back-transformation to the solutions of the first order equation. In the paper \cite{nowakowski_1998} from heuristic considerations such a higher (third) order wave equation possessing a necessary property of the reversibility was proposed. However, by virtue of that the higher order equation does not reduce to the Klein-Gordon-Fock equation in the interaction free case, this leads to the delicate question of physical interpretation of the terms in such a higher order equation.\\
\indent In our paper \cite{markov_2015} this approach was analyzed in more detail. We have suggested a scheme of systematic deriving the wave equation of third order and obtained the most general form of this equation in comparison with a similar equation in the paper by Nowakowski \cite{nowakowski_1998}. This scheme enables one in principle to obtain the wave equations of higher order in derivatives for a description of particles with a spin greater than 1 (the case of $s = 3/2$ was discussed in \cite{markov_2017}).\\
\indent We have established that the construction of the required divisor $d(D)$, which would commute with the $L(D)$-operator, is closely related with a problem of constructing a cubic root of  the third order (massive) wave operator in the interaction free case. By a direct calculation we have shown that by using only the algebra of Duffin-Kemmer-Petiau matrices, it is impossible to calculate the required cubic root and thereby eventually to calculate the required divisor $d(D)$. For solving this problem we had to introduce into consideration an additional algebraic object, the so-called $q$-commutator ($q$ is a deformation parameter, representing a primitive cubic root of unity) and a new set of matrices $\eta_{\mu}(z)$ instead of the original $\beta_{\mu}$-matrices of the DKP-algebra. In a general case these matrices depend on an arbitrary complex parameter $z$ and Schr\"odinger's ``pseudomatrix'' $\omega$ and are not connected by any unitary transformation with the $\beta_{\mu}$-matrices. We have shown that based on new algebraic objects a procedure of constructing cubic root of the third order wave operator can be reduced to a few simple algebraic transformations and an operation of the passage to the limit $z \rightarrow q$. In other words, the third order wave operator (without interaction) is obtained as a finite limit of the cube of some first order differential operator $\hat{\cal L}(z,D)$. The latter is singular at $z = q$. The definitions of this operator, of the matrices $\eta_{\mu}(z)$, and of the pseudomatrix $\omega$ will be given just below.\\
\indent We have made corresponding generalization of the result obtained to the case of the presence of an external electromagnetic field in the system and performed a detail comparison with the result of Nowakowski. This gives us the possibility to have a new way of looking at the problem of constructing the propagator of a massive vector particle in an external gauge field in the form of path integral in parasuperspace within the framework of Duffin-Kemmer-Petiau theory with the deformation. As discussed above, the lack of commutativity of the Klein-Gordon-Fock divisor in the case of spin-1 particle with the original DKP-operator $L(D)$ in the presence of a gauge field in the system leads to that we can not define the Fock-Schwinger proper-time representation for the inverse DKP-operator $L^{-1}(D)$, i.e. already at the very first step of constructing the desired integral representation we are faced with the problem of a fundamental character. We can overcome this difficulty only by redefining the original DKP-operator $L(D)$ and corresponding divisor $d(D)$.\\
\indent This a rather drastic step has allowed us \cite{markov_2015} to write almost immediately the Fock-Schwinger proper-time representation for the inverse operator $\hat{\cal L}^{-1}(z)$:
\begin{equation}
\frac{1}{\hat{\cal L}(z)} \equiv \frac{\hat{\cal L}^{2}(z)}{\hat{\cal L}^{3}(z)} =
-i\!\int\limits_{0}^{\infty}\!d\hspace{0.03cm}T\!
\int\!\frac{d^{\,2}\chi}{T^{2}}\;\hspace{0.02cm}
{\rm e}^{\displaystyle{-i\hspace{0.03cm}T\bigl (\hat{H}(z) - i\hspace{0.02cm}\epsilon\hspace{0.02cm}\bigr)
+
\frac{1}{2}\,\bigl(\hspace{0.02cm}T\hspace{0.02cm}[\hspace{0.03cm}\chi,
\hat{\cal L}(z)\hspace{0.02cm}]
+
\frac{1}{4}\,T^{2\,}[\hspace{0.03cm}\chi,
\hat{\cal L}(z)\hspace{0.02cm}]^{\hspace{0.02cm}2}\hspace{0.03cm}\bigr)}}\!\!,
\;
\epsilon\rightarrow +\hspace{0.01cm}0,
\label{eq:2w}
\end{equation}
where
\begin{equation}
\hat{\cal L}(z) \equiv \hat{\cal L}(z,D) =
A\hspace{0.02cm}\biggl(\frac{\!i}{\,\varepsilon^{1/3}(z)}\,\eta_{\mu}(z)
\hspace{0.03cm}D_{\mu} + m\hspace{0.02cm}I\biggr)
\label{eq:2e}
\end{equation}
and
\begin{equation}
\hat{H}(z) \equiv \hat{\cal L}^{\hspace{0.02cm}3}(z)
\label{eq:2r}
\end{equation}
is the Hamilton operator, $D_{\mu} = \partial_{\mu} + ieA_{\mu}(x)$ is the covariant derivative. The Greek letters $\mu, \nu,\ldots$ run from 1 to $2\hspace{0.01cm}M$  unless otherwise stated, and $\chi$ is a para-Grassmann variable of order $p = 2$ (i.e. $\chi^{3} = 0$) with the rules of an integration \cite{omote_1979}:
\[
\int\!d^{\,2}\chi  = 0,
\quad
\int\!d^{\,2}\chi\,[\hspace{0.03cm}\chi,\hat{\cal L}\hspace{0.03cm}] = 0,
\quad
\int\!d^{\,2}\chi\,[\hspace{0.03cm}\chi,\hat{\cal L}\hspace{0.03cm}]^{\,2}
=
4\hspace{0.03cm}i^{\hspace{0.03cm}2\!}\hat{\cal L}^{\hspace{0.02cm}2}.
\]
In (\ref{eq:2e}) we have introduced the function
\[
\varepsilon\hspace{0.01cm}(z) = 1 + z + z^{2} \equiv (z - q)(z - q^{2}),
\]
where $q$ is a primitive cubic root of unity. As a proper para-supertime it is necessary to take a triple $(T, \chi, \chi^2)$. Note that the representation (\ref{eq:2w}) implicitly supposes the validity of the following relations:
\begin{equation}
[\hspace{0.03cm}\hat{H}(z), [\hspace{0.03cm}\chi,
\hat{\cal L}(z)\hspace{0.04cm}]\hspace{0.03cm}] = 0,
\qquad
[\hspace{0.02cm}\chi,\hat{\cal L}(z)\hspace{0.04cm}]^{\hspace{0.02cm}3} = 0.
\label{eq:2t}
\end{equation}
It is far less trivial to prove (\ref{eq:2t}) and really it is a good test to check the self-consistency of the approach under consideration as a whole\footnote{\,In fact an analysis of the relations of the type (\ref{eq:2t}) even in the case of spin $1/2$ in the presence of an external electromagnetic field is not quite simple and this delicate point for some reason is not discussed at all in literature (see, for example, \cite{fradkin(1)_1991}). Here, instead of (\ref{eq:2e}) and (\ref{eq:2r}) we have
\begin{equation}
 \hat{\cal L}(D) = \gamma_{5} \hspace{0.02cm}(\hspace{0.02cm}
i\hspace{0.02cm}\gamma_{\mu}\hspace{0.02cm}D_{\mu} + m\hspace{0.02cm}I)
\quad \mbox{and} \quad
\hat{H} \equiv \hat{\cal L}^{\hspace{0.02cm}2}.
\label{eq:2y}
\end{equation}
The reason of complication in the analysis of the first relation in (\ref{eq:2t}) is that, for example, in the operator realization of the Dirac-Clifford algebra in terms of Grassmann variables and their derivatives the operators $\hat{\gamma}_{\mu}$ are Grassmann-odd (fermionic) operators while the realization of $\hat{\gamma}_{5} \equiv - ({1}/{4!})\hspace{0.04cm}\epsilon_{\mu \nu \lambda \sigma}\hat{\gamma}_{\mu} \hat{\gamma}_{\nu}\hat{\gamma}_{\lambda} \hat{\gamma}_{\sigma}$ results in a Grassmann-even (bosonic) operator. Van Holten in the paper \cite{holten_1995} was the first to point out this fact of mixing the terms with different Grassmann parity by a non-zero mass term in (\ref{eq:2y}). It is precisely this circumstance that leads the first relation in (\ref{eq:2t}) to  require the Maxwell background field to satisfy equation of motion. In the case of a spin-1 particles the situation becomes more entangled. We will consider all these points in our subsequent paper \cite{part_II}, when mathematical technique required for this purpose will be developed.}. The operator $\hat{\cal L}(z, D)$ represents the cubic root of some third order wave operator in an external electromagnetic field. Matrix element of the inverse operator $\hat{\cal L}^{-1}(z, D)$ in the corresponding basis of states can be considered as a propagator of a massive vector particle in the background gauge field (see the next section).\\
\indent Further, the matrices $\eta_{\mu}(z)$ are defined by the matrices $\beta_{\mu}$ obeying the Duffin-Kemmer-Petiau algebra (\ref{eq:2q}) and by the complex deformation parameter $z$ as follows:
\begin{equation}
\eta_{\mu}(z) = \biggl(1 + \frac{1}{2}\,z\biggr)\beta_{\mu} + z\hspace{0.03cm}\biggl(\frac{i\sqrt{3}}{2}\biggr)\hspace{0.03cm}
[\hspace{0.03cm}\omega, \beta_{\mu}\hspace{0.03cm}],
\label{eq:2u}
\end{equation}
where
\begin{equation}
\omega = \frac{\!\!1}{(M!)^{\hspace{0.02cm}2}}\; \epsilon_{\mu_{1}\mu_{2}\ldots\hspace{0.02cm}\mu_{2M}}
\beta_{\mu_{1}}\beta_{\mu_{2}}\ldots\beta_{\mu_{2M}}.
\label{eq:2i}
\end{equation}
 In view of the definition of the $\omega$ matrix, Eq.\,(\ref{eq:2i}), and of the trilinear relation for the $\beta$-matrices, Eq.\,(\ref{eq:2q}), we have an important property \cite{harish-chandra_1947, harish-chandra_1946, fujiwara_1953}
\begin{equation}
\omega^{3} = \omega.
\label{eq:2o}
\end{equation}
Let us note only that the matrix $\omega$ is identically zero for the spin 0 (five-dimensional irreducible representation of the DKP algebra in the four-dimensional Euclidean space-time). Therefore, only the ten-row representation for the spin-1 case needs to be considered.\\
\indent The matrix $A$ in the expression (\ref{eq:2e}) was determined by us \cite{markov_2015} in the form of the expansion in powers of $\omega$:
\begin{equation}
A = \alpha\hspace{0.02cm}I  + \beta\hspace{0.02cm}\omega
+ \gamma\hspace{0.04cm}\omega^{2} 
\equiv
\alpha\hspace{0.04cm}{\rm e}^{\textstyle i\hspace{0.03cm}\frac{2\hspace{0.02cm}\pi}{3}\,\omega},
\label{eq:2p}
\end{equation}
where the coefficients are
\[
\beta = \biggl(\frac{i\sqrt{3}}{2}\biggr)\hspace{0.02cm}\alpha,
\quad
\gamma = \biggl(-\frac{3}{2}\biggr)\hspace{0.02cm}\alpha,
\quad
\alpha^{3} = \frac{1}{m}\,,
\]
and $I$ is the unit matrix. For the exponential representation in (\ref{eq:2p}) the property (\ref{eq:2o}) was taken into account.\\ 
\indent At the end of all calculations, it should be necessary to proceed to the limit $z\rightarrow q$ and in particular, in this limit the operator $\hat{H}(z)$, Eq.\,(\ref{eq:2r}), defines the third-order wave operator in an external electromagnetic field
\begin{align}
\hat{H} &= \lim_{z\hspace{0.02cm}\rightarrow\hspace{0.03cm} q}\hspace{0.02cm} \hat{H}(z)  \notag\\
&= \lim_{z\hspace{0.02cm}\rightarrow\hspace{0.03cm} q}\hspace{0.02cm} \hat{\cal L}^{3}(z,D) =
\lim_{z\hspace{0.02cm}\rightarrow\hspace{0.03cm} q}\hspace{0.02cm}
\biggl[\hspace{0.03cm}A\hspace{0.02cm}\biggl(\frac{\!i}{\,\varepsilon^{1/3}(z)}\,\eta_{\mu}(z)
\hspace{0.03cm}D_{\mu} + m\hspace{0.02cm}I\biggr)\biggr]^{3}.
\notag
\end{align}
An explicit form of this limit is given in the paper \cite{markov_2015}.\\
\indent We note that the argument of the exponential in the Fock-Schwinger proper-time representation (\ref{eq:2w}) is in a good agreement with the structure of the action for a relativistic classical spin-1 particle (\ref{eq:1q}), defined in terms of para-Grassmann variables. However, a kinetic part of the action (\ref{eq:1q}) was chosen in a complete analogy with the kinetic parts of classical and quantum models of Dirac's particle, whereas we expect based on a general formula of the representation (\ref{eq:2w}) that the situation here can be more complicated since the operator $\hat{H}(z)$ contains the third order derivatives with respect to $x_{\mu}$. We adopt the Fock-Schwinger representation (\ref{eq:2w}) for the inverse operator ${\cal L}^{-1}(z,D)$ with the deformation as an initial expression for constructing representation in the form of path integral with the use of corresponding system of coherent states in a close analogy with the paper by Borisov and Kulish \cite{borisov_1982} for the case of spin 1/2.


\section{\bf The operator formalism}
\setcounter{equation}{0}
\label{section_3}

The starting point of our study is the Fock-Schwinger proper-time representation (\ref{eq:2w}). The problem of finding the Green's function ${\cal D}_{\alpha \beta}(x^{\prime}, x; z)$ of a massive vector particle in an external electromagnetic field
\[
\biggl[A\hspace{0.02cm}\biggl(\frac{\!i}{\,\varepsilon^{1/3}(z)}\,\eta_{\mu}(z)
\hspace{0.03cm}D_{\mu} + m\hspace{0.02cm}I\biggr)\biggr]_{\alpha\gamma}
{\cal D}_{\gamma\beta}(x^{\prime}\!,x\hspace{0.02cm};z)
=
\delta_{\alpha\beta}\hspace{0.02cm}\delta(x^{\prime} - x),
\]
reduces to the construction of an operator that is the inverse of the operator
\begin{equation}
\hat{\cal L}(z) \equiv \hat{\cal L}(z,\hat{D}) =
\hat{A}\hspace{0.03cm}\biggl(\frac{\!i}{\,\varepsilon^{1/3}(z)}\,\hat{\eta}_{\mu}(z)
\hspace{0.03cm}\hat{D}_{\mu} + m\hspace{0.02cm}\hat{I}\biggr),
\label{eq:3q}
\end{equation}
where $ \mu = 1,2,\,\ldots\,,2M$; $\alpha,\,\beta,\,\gamma = 1,2,\,\ldots\,,n^{(2M)}_{M}$ and $n^{(2M)}_{M}\! = C^{2M + 1}_{M}$ is the highest rank of the irreducible representations of the DKP algebra with an even number $2M$ of the elements $\beta_{\mu}$. Hereinafter, we use the notation of quantities with hat above for those operators, which need to be distinguished from their matrix analogue. We restrict our consideration to the most important case $M = 2$ that corresponds to the four-dimension Euclidean space-time.\\
\indent The operator $\hat{\cal L}^{-1}(z, \hat{D})$ acts on the space ${\cal H}$ of the representation of the algebras
\begin{equation}
[\hspace{0.02cm}\hat{p}_{\mu},\hat{x}_{\nu}\hspace{0.02cm}] = i\hspace{0.02cm}\delta_{\mu\nu},
\label{eq:3w}
\end{equation}
\begin{equation}
\hat{\beta}_{\mu}\hat{\beta}_{\nu}\hat{\beta}_{\lambda} + \hat{\beta}_{\lambda}\hat{\beta}_{\nu}\hat{\beta}_{\mu}
=
\delta_{\mu\nu}\hat{\beta}_{\lambda} + \delta_{\lambda\nu}\hat{\beta}_{\mu}.
\label{eq:3e}
\end{equation}
The space ${\cal H}$ is determined in the form of the tensor product of two spaces $H$ and $L$, which realize representations of each algebra (\ref{eq:3w}) and (\ref{eq:3e}). The Green's function ${\cal D} (x^{\prime}, x; z)$ is a matrix element of the operator $\hat{\cal L}^{-1}(z, \hat{D})$ in the basis $\{\vert\,x\hspace{0.02cm}\rangle;\, x\in R^{4}\}$ in $H$ and in the matrix basis $\{ \vert\,\alpha\hspace{0.02cm}\rangle;\, \alpha =1, 2, \ldots , 10\}$ in $L$:
\[
{\cal D}_{\alpha\beta}(x^{\prime}\!,x\hspace{0.02cm};z)
=
\langle\hspace{0.02cm}x^{\prime},\alpha\hspace{0.02cm}|\,
\hat{\cal L}^{-1}(z,\hat{D})\hspace{0.01cm}|\hspace{0.03cm}
x,\beta\hspace{0.02cm}\rangle.
\]
\indent To construct the path integral, we will need a basis of coherent states in the spin-1 space $L$. In $L$, the representation space of the Duffin-Kemmer-Petiau operator algebra (\ref{eq:3e}), in accordance with (\ref{ap:A2}) we introduce the parafermion creation and annihilation operators
\begin{equation}
a^{\pm}_{1} = \hat{\beta}^{\phantom{\pm}\!}_{1} \pm i\hspace{0.02cm}\hat{\beta}^{\phantom{\pm}\!}_{2},
\qquad
a^{\pm}_{2} = \hat{\beta}^{\phantom{\pm}\!}_{3} \pm i\hspace{0.02cm}\hat{\beta}^{\phantom{\pm}\!}_{4}.
\label{eq:3r}
\end{equation}
These operators by virtue of (\ref{eq:3e}) obey the following algebra:
\begin{align}
&a^{\pm}_{k}a^{\mp}_{l}a^{\pm}_{m} + a^{\pm}_{m}a^{\mp}_{l}a^{\pm}_{k}
=
2\hspace{0.02cm}\delta^{\phantom{\pm}\!}_{kl}\hspace{0.04cm}a^{\pm}_{m}
+
2\hspace{0.02cm}\delta^{\phantom{\pm}\!}_{ml}\hspace{0.03cm}a^{\pm}_{k},
\label{eq:3t} \\[1ex]
&a^{\pm}_{k}a^{\mp}_{l}a^{\mp}_{m} + a^{\mp}_{m}a^{\mp}_{l}a^{\pm}_{k}
=
2\hspace{0.02cm}\delta^{\phantom{\pm}\!}_{kl}\hspace{0.04cm}a^{\mp}_{m},
\label{eq:3y} \\[1ex]
&a^{\pm}_{k}a^{\pm}_{l}a^{\pm}_{m} + a^{\pm}_{m}a^{\pm}_{l}a^{\pm}_{k}
= 0,\quad
k,l,m = 1, 2
\label{eq:3u}
\end{align}
and the space $L$ can be realized as a finite Fock space for the para-Fermi operators $(a_{1}^{\pm}, a_{2}^{\pm})$.\\
\indent As coherent states of the para-Fermi operators we take the coherent states as they were defined by Omote and Kamefuchi \cite{omote_1979}. For papastatistics $p = 2$ they have the form (in the case when $M = 2$):
\begin{equation}
\begin{split}
&|\hspace{0.02cm}(\xi)_{2}\hspace{0.02cm}\rangle = \exp\Bigl(-\frac{1}{2}\sum^{2}_{l\hspace{0.02cm} =
\hspace{0.02cm}1}\,
[\hspace{0.04cm}\xi^{\phantom{+\!\!}}_{l}, a^{+}_{l}\hspace{0.02cm}]\Bigr)|\hspace{0.03cm}0\rangle,
\\[1ex]
&\langle\hspace{0.02cm}(\bar{\xi}^{\,\prime})_{2}\hspace{0.02cm}|
=
\langle\hspace{0.03cm}0|\exp\Bigl(\hspace{0.02cm}\frac{1}{2}\sum^{2}_{l=1}\,
[\hspace{0.04cm}\bar{\xi}^{\,\prime}_{l}, a^{\!-}_{l}\hspace{0.03cm}]\Bigr),
\end{split}
\label{eq:3i}
\end{equation}
so that
\[
a^{\!-}_{k}|\hspace{0.04cm}(\xi)_{2}\hspace{0.02cm}\rangle = \xi^{\phantom{-\!\!}}_{k}|\hspace{0.04cm}(\xi)_{2}\hspace{0.02cm}\rangle,
\qquad
\langle\hspace{0.02cm}(\bar{\xi}^{\,\prime})_{2}\hspace{0.02cm}|\,a^{+}_{k}
=
\langle\hspace{0.02cm}(\bar{\xi}^{\,\prime})_{2}\hspace{0.02cm}|\, \bar{\xi}^{\,\prime}_{k},
\]
where $\xi^{\phantom{\prime\!\!}}_{k}$ and $\bar{\xi}_{k}^{\prime},\, k=1,2$ are para-Grassmann numbers obeying algebra (\ref{ap:B2}). For brevity sometimes we will write
\[
\sum^{2}_{l\hspace{0.02cm}=
\hspace{0.02cm}1}\;
[\hspace{0.04cm}\xi^{\phantom{+\!\!}}_{l}, a^{+}_{l}\hspace{0.02cm}]
\equiv
[\hspace{0.04cm}\xi\hspace{0.03cm}, a^{+}\hspace{0.02cm}],
\qquad
\sum^{2}_{l\hspace{0.02cm}=
\hspace{0.02cm}1}\;
[\hspace{0.04cm}\bar{\xi}^{\,\prime}_{l}, \hspace{0.03cm}\xi^{\phantom{\prime}}_{l}\hspace{0.03cm}]
\equiv
[\hspace{0.04cm}\bar{\xi}^{\,\prime}, \hspace{0.03cm}\xi\hspace{0.03cm}]
\]
and moreover since we are interested in only the case parastatistics of order 2, then we will omit the symbol 2 in the notation of the parafermion coherent states, i.e.
\[
|\hspace{0.03cm}(\xi)_{2}\hspace{0.02cm}\rangle
\equiv
|\,\xi\hspace{0.02cm}\rangle,
\qquad
\langle\hspace{0.02cm}(\bar{\xi}^{\,\prime})_{2}\hspace{0.01cm}|
\equiv
\langle\hspace{0.02cm}\bar{\xi}^{\,\prime}\hspace{0.02cm}|.
\]
The overlap function and completeness relation for the coherent states (\ref{eq:3i}) are given by
\[
\langle\hspace{0.025cm}\bar{\xi}^{\,\prime}\hspace{0.02cm}
|\,\xi\hspace{0.02cm}\rangle
=
\exp\Bigl\{\frac{1}{2}\,
[\hspace{0.02cm}\bar{\xi}^{\,\prime},\xi\hspace{0.02cm}\hspace{0.02cm}]\!\hspace{0.02cm}
\Bigr\},
\]
\[
\iint\!|\,\xi\hspace{0.02cm}\rangle\hspace{0.02cm} \langle\hspace{0.02cm}\bar{\xi}\hspace{0.02cm}|\,
{\rm e}^{\,-\textstyle\frac{\!1}{2}\,[\hspace{0.03cm}\bar{\xi},\xi\hspace{0.03cm}]}
\hspace{0.03cm}(d\xi)_{2}\hspace{0.03cm}(d\bar{\xi})_{2} = \hat{1},
\]
where
\[
(d\xi)_{2}\equiv d^{\hspace{0.02cm}2}\xi_{2}\hspace{0.03cm} d^{\hspace{0.02cm}2}\xi_{1}, \quad
(d\bar{\xi})_{2}\equiv d^{\hspace{0.02cm}2}\bar{\xi}_{1}\hspace{0.03cm} d^{\hspace{0.02cm}2}\bar{\xi}_{2}.
\]
\indent The transition from the matrix elements in the coherent basis to the representation in which the DKP matrices 
$\beta_{\mu}$ have a specific form is realized as follows:
\[
\langle\hspace{0.02cm}\alpha\hspace{0.02cm}|\hspace{0.03cm}\ldots\hspace{0.01cm}
|\hspace{0.04cm}\beta\hspace{0.02cm}\rangle
=\!
\iint\!{\rm e}^{\,-\textstyle\frac{\!1}{2}\,[\hspace{0.03cm}\bar{\xi}^{\prime},\xi^{\prime}\hspace{0.03cm}]}
\hspace{0.03cm}(d\xi^{\prime})_{2}\hspace{0.03cm}(d\bar{\xi}^{\prime})_{2}
\,
{\rm e}^{\,-\textstyle\frac{\!1}{2}\,[\hspace{0.03cm}\bar{\xi},\xi\hspace{0.03cm}]}
\hspace{0.03cm}(d\xi)_{2}\hspace{0.03cm}(d\bar{\xi})_{2}
\,
\langle\hspace{0.02cm}\alpha\hspace{0.02cm}|\,\xi^{\prime}\hspace{0.02cm}\rangle
\hspace{0.03cm}\langle\hspace{0.025cm}\bar{\xi}^{\,\prime}\hspace{0.02cm}|\ldots
|\,\xi\hspace{0.02cm}\rangle\hspace{0.03cm}
\langle\hspace{0.025cm}\bar{\xi}\hspace{0.02cm}|\hspace{0.04cm}\beta\hspace{0.03cm}\rangle.
\]
The calculation of the explicit form of the transition functions $\langle\hspace{0.02cm}\alpha\hspace{0.02cm}|\,\xi\hspace{0.02cm}\rangle$ and $\langle\hspace{0.025cm}\bar{\xi}\hspace{0.02cm}|\hspace{0.04cm}\beta\hspace{0.03cm}\rangle$ will be considered in Part II \cite{part_II}.\\
\indent To present the propagator ${\cal D}_{\alpha\beta}(x^{\prime}, x; z)$ in the form of a path integral in parasuperspace of an exponential whose argument is the classical action for the massive vector particle, we use the operator formalism and the Fock-Schwinger proper time representation for the inverse operator $\hat{\cal L}^{-1}(z)$, Eq.\,(\ref{eq:2w}). We rewrite the matrix element of the inverse operator $\hat{\cal L}^{-1}(z)$ in the form
\begin{equation}
\langle\hspace{0.02cm}x^{\prime},\bar{\xi}^{\,\prime}\hspace{0.02cm}|\,
\frac{1}{\hat{\cal L}(z)}\,|\hspace{0.04cm}x,\xi\hspace{0.02cm}\rangle
\equiv
\langle\hspace{0.02cm}x^{\prime},\bar{\xi}^{\,\prime}\hspace{0.02cm}|\,
\frac{\hat{\cal L}^{2}(z)}{\hat{\cal L}^{3}(z)}\,|\hspace{0.04cm}
x,\xi\hspace{0.02cm}\rangle
=
\label{eq:3o}
\end{equation}
\[
=-i\!\int\limits_{0}^{\infty}\!d\hspace{0.03cm}T\!
\int\!\frac{d^{\,2}\chi}{T^{2}}\;\hspace{0.02cm}
\langle\hspace{0.02cm}x^{\prime},\bar{\xi}^{\,\prime}\hspace{0.02cm}|\;
{\rm e}^{\displaystyle{-\hspace{0.02cm}i\hspace{0.03cm}T\bigl (\hat{H}(z) - i\hspace{0.02cm}\epsilon\hspace{0.02cm}\bigr)
+
\frac{1}{2}\,\bigl(\hspace{0.02cm}T\hspace{0.02cm}
[\hspace{0.03cm}\chi,\hat{\cal L}(z)\hspace{0.03cm}]
+
\frac{1}{4}\,T^{2\,}[\hspace{0.03cm}\chi,
\hat{\cal L}(z)\hspace{0.03cm}]^{\hspace{0.02cm}2}\hspace{0.03cm}\bigr)}}
|\hspace{0.04cm}x,\xi\hspace{0.02cm}\rangle,
\quad
\epsilon\rightarrow +\hspace{0.01cm}0.
\]
\indent Further, in accordance with Tobocman \cite{tobocman_1956}, we have to divide the interval $[0, T]$ into $N$ parts, $T = \Delta\tau N$ and to represent the exponential in matrix element (\ref{eq:3o}) in the form of a product of $N$ exponential multiplies
\[
{\rm e}^{\displaystyle{-\hspace{0.02cm}i\hspace{0.03cm}T\hspace{0.03cm}\hat{H}(z)
+
\frac{1}{2}\,T\hspace{0.02cm}
[\hspace{0.03cm}\chi,\hat{\cal L}(z)\hspace{0.03cm}]
+\, \ldots}}
=
\biggl({\rm e}^{\displaystyle{-\hspace{0.02cm}i\hspace{0.03cm}\Delta\tau\hat{H}(z)
+
\frac{1}{2}\,\Delta\tau\hspace{0.02cm}
[\hspace{0.03cm}\chi,\hat{\cal L}(z)\hspace{0.03cm}]
+\,\ldots\,}}\biggr)^{\!\!N}.
\]
Such a factorization of the exponential is well defined for the part linear in $T$. However, in the exponential we have also the term quadratic in $T$ resulting in qualitative difference from the standard consideration. Let us analyze this important point in more detail.\\
\indent We introduce a generalized Hamilton operator explicitly depending from ``time'' $\tau$:
\begin{equation}
\hat{\cal H}(\tau;z) = \hat{H}(z) +
\frac{1}{2}\,[\hspace{0.03cm}\chi,\hat{\cal L}(z)\hspace{0.01cm}]
+
\frac{1}{4}\,\tau\,[\hspace{0.03cm}\chi,\hat{\cal L}(z)\hspace{0.01cm}]^{\hspace{0.02cm}2},
\quad 0\leq\tau\leq T.
\label{eq:3p}
\end{equation}
In the paper by Mizrahi \cite{mizrahi_1975} the problem of path integral representation for a system in which a Hamiltonian explicitly depends on time, was considered. Here, we will follow the approach presented in this work.\\
\indent For the construction of the required representation it is necessary to ensure that the following condition holds:
\begin{equation}
[\hspace{0.04cm}\hat{\cal H}(\tau;z),\hat{\cal H}(s;z)\hspace{0.02cm}] = 0,
\quad
\tau,\hspace{0.02cm} s\in [\hspace{0.03cm}0,T\hspace{0.03cm}].
\label{eq:3a}
\end{equation}
By virtue of the definition (\ref{eq:3p}) this requirement reduces to the first relation in (\ref{eq:2t}). For further formalization of the task it is convenient to define an evolution operator as follows:
\[
\hat{U}\hspace{0.02cm}(T, 0) = {\rm e}^{\displaystyle-i\! \int_{0}^{T}\! ds\:
\hat{\cal H}(s\hspace{0.03cm};z)}.
\]
The condition (\ref{eq:3a}) assures a correctness of the following decomposition:
\begin{equation}
\hat{U}\hspace{0.02cm}(T, 0)
=
\hat{U}\hspace{0.02cm}(\tau_{N}, \tau_{N - 1})\hspace{0.02cm}\hat{U}\hspace{0.02cm}
(\tau_{N - 1}, \tau_{N - 2})
\,\ldots\,\hat{U}\hspace{0.02cm}(\tau_{1}, \tau_{0}),
\label{eq:3s}
\end{equation}
where $\tau_N \equiv N,\, \tau_0=0$ and
\[
\hat{U}\hspace{0.02cm}(\tau_{j}, \tau_{j - 1}) =
{\rm e}^{\displaystyle-i\! \int_{\tau_{j - 1}}^{\tau_{j}}\! ds\: \hat{\cal H}(s\hspace{0.03cm};z)}.
\]
For the last expression, in view of (\ref{eq:3p}), we have
\begin{equation}
\hat{U}\hspace{0.02cm}(\tau_{j}, \tau_{j - 1})
=
{\rm e}^{\displaystyle{-\hspace{0.02cm}i\hspace{0.03cm}\Delta\tau\bigl\{\hat{H}(z)
+
\frac{1}{2}\,[\hspace{0.03cm}\chi,\hat{\cal L}(z)\hspace{0.03cm}]
+
\frac{1}{8}\,(\tau_{j} + \tau_{j - 1})\hspace{0.03cm}
[\hspace{0.03cm}\chi,\hat{\cal L}(z)\hspace{0.03cm}]^{\hspace{0.03cm}2}
\hspace{0.02cm}\bigr\}}}.
\label{eq:3d}
\end{equation}
In the limit $N\rightarrow\infty,\,\Delta\tau\rightarrow 0$ it should be considered that
\[
\tau_{j} + \tau_{j - 1} \rightarrow 2\hspace{0.03cm}\tau,
\]
i.e. an effective Lagrangian in the classical action for the massive vector particle will depend on the additional continuous parameter $\tau$. Thus, instead of the standard decomposition \cite{tobocman_1956}
\[
{\rm e}^{\displaystyle{-\hspace{0.02cm}i\hspace{0.03cm}T\hspace{0.02cm}\hat{H}}}
=
\underbrace{
{\rm e}^{\displaystyle{-\hspace{0.02cm}i\hspace{0.03cm}\Delta\tau\hspace{0.03cm}\hat{H}}}
{\rm e}^{\displaystyle{-\hspace{0.02cm}i\hspace{0.03cm}\Delta\tau\hspace{0.03cm}\hat{H}}}
\,\ldots\,
{\rm e}^{\displaystyle{-\hspace{0.02cm}i\hspace{0.03cm}\Delta\tau\hspace{0.03cm}\hat{H}}}
}_{N\, \mbox{\footnotesize times}}\;,
\quad 
\Delta\tau\hspace{0.03cm}N = T,
\]
in our case we will use the decomposition (\ref{eq:3s}) with (\ref{eq:3d}) and insert  resolutions of the identity in $H\otimes L$ between the evolution operators $\hat{U}(\tau_{j}, \tau_{j-1})$. Following Borisov and Kulish \cite{borisov_1982} in the $k$-th position, we insert
\[
\hat{I}_{x} =
\int\prod\limits^{4}_{\mu\hspace{0.02cm} = \hspace{0.02cm}1}dx^{(k)}_{\mu}\!\!
\iint\!
{\rm e}^{\,-\textstyle\frac{\!1}{2}\,[\hspace{0.03cm}\bar{\xi}^{(k)},\xi^{(k)}\hspace{0.03cm}]}
\hspace{0.03cm}(d\xi^{(k)})_{2}\hspace{0.03cm}(d\bar{\xi}^{(k)})_{2}\hspace{0.02cm}
|\,x^{(k)},\xi^{(k)}\hspace{0.02cm}\rangle\hspace{0.02cm}
\langle\hspace{0.02cm}x^{(k)}, \bar{\xi}^{(k)}\hspace{0.02cm}|\hspace{0.02cm}.
\]
\indent Since the evolution operator $\hat{U}(\tau_{j}, \tau_{j-1})$ contains the noncommuting operators $\hat{p}_{\mu},\,\hat{x}_{\mu},\,a^{\pm}_{n}$, for obtaining the explicit form of the matrix elements
\[
\langle\hspace{0.02cm}(k)_{x}\hspace{0.02cm}|\hspace{0.02cm}
\hat{U}\hspace{0.02cm}(\tau_{k}, \tau_{k - 1}) \hspace{0.01cm}|
\hspace{0.02cm}(k - 1)_{x}\hspace{0.02cm}\rangle
\equiv
\langle\hspace{0.02cm}x^{(k)}, \bar{\xi}^{(k)}\hspace{0.02cm}|
\hspace{0.02cm}
\hat{U}\hspace{0.02cm}(\tau_{k}, \tau_{k - 1}) \hspace{0.01cm}
|\,x^{(k- 1)},\xi^{(k - 1)}\hspace{0.02cm}\rangle
\]
it is necessary to use an additional resolution of the identity:
\[
\hat{I}_{p} =
\int\prod\limits^{4}_{\mu\hspace{0.02cm} = \hspace{0.02cm}1}dp^{(k)}_{\mu}\!\!
\iint\!
{\rm e}^{\,-\textstyle\frac{\!1}{2}\,
[\hspace{0.03cm}\bar{\xi}^{\hspace{0.02cm}\prime(k)},
\xi^{\hspace{0.02cm}\prime(k)}\hspace{0.03cm}]}
\hspace{0.03cm}(d\xi^{\hspace{0.02cm}\prime(k)})_{2}\hspace{0.03cm}
(d\bar{\xi}^{\hspace{0.02cm}\prime(k)})_{2}\hspace{0.02cm}
|\,p^{(k)},\xi^{\hspace{0.02cm}\prime(k)}\hspace{0.02cm}\rangle\hspace{0.02cm}
\langle\hspace{0.02cm}p^{(k)}, \bar{\xi}^{\hspace{0.02cm}\prime(k)}\hspace{0.02cm}|\hspace{0.02cm}.
\]
Thus the matrix element of evolution operator $\hat{U}(T, 0)$ takes the form:
\[
\langle\hspace{0.02cm}x^{\prime}, \bar{\xi}^{\prime}\hspace{0.02cm}|
\hspace{0.04cm}\hat{U}\hspace{0.02cm}(T, 0) \hspace{0.01cm}
|\,x,\xi\hspace{0.02cm}\rangle
=
\vspace{0.2cm}
\]
\[
\langle x^{\prime}, \bar{\xi}^{\prime}|
\hspace{0.02cm}\hat{I}^{(N)}_{x}\hat{I}^{(N)}_{p}
\hat{U}(\tau_{N},\!\tau_{N\!- 1})\hspace{0.02cm}
\hat{I}^{(N\! - 1)}_{x}\hat{I}^{(N\! - 1)}_{p}
\hat{U}(\tau_{N\!- 1},\!\tau_{N\!- 2})
\hat{I}^{(N\!- 2)}_{x}\hat{I}^{(N\!- 2)}_{p}
\!\ldots
\hat{U}(\tau_{2},\!\tau_{1})
\hat{I}^{(1)}_{x}\hat{I}^{(1)}_{p}
\hat{U}(\tau_{1},\!\tau_{0})
|\hspace{0.02cm}x,\xi\rangle
\]
and the following analysis, in view of (\ref{eq:3d}), reduces to the calculation of the matrix element
\begin{equation}
\langle\hspace{0.02cm}(k)^{\prime}_{p}\hspace{0.02cm}|\hspace{0.02cm}
\hat{U}\hspace{0.02cm}(\tau_{k}, \tau_{k - 1}) \hspace{0.01cm}|\hspace{0.02cm}
(k - 1)_{x}\hspace{0.02cm}\rangle
\simeq
\label{eq:3g}
\end{equation}
\[
\simeq
\langle\hspace{0.02cm}(k)^{\prime}_{p}\hspace{0.02cm}|\hspace{0.02cm}
1 - i\hspace{0.03cm}\Delta\tau\bigl\{\hat{H}(z)
+
\frac{1}{2}\,[\hspace{0.03cm}\chi,\hat{\cal L}(z)\hspace{0.03cm}]
+
\frac{1}{8}\,(\tau_{k} + \tau_{k - 1})\hspace{0.03cm}
[\hspace{0.03cm}\chi,\hat{\cal L}(z)\hspace{0.03cm}]^{\hspace{0.03cm}2}
\hspace{0.02cm}\bigr\}
|\hspace{0.02cm}(k - 1)_{x}\hspace{0.02cm}\rangle
\]
with the overlap function
\[
\langle\hspace{0.02cm}(k)^{\prime}_{p}\hspace{0.03cm}|
\hspace{0.03cm}(k - 1)_{x}\hspace{0.02cm}\rangle
=
\frac{1}{(2\pi)^{2}}\,
\exp\hspace{0.02cm}\biggl\{i\hspace{0.02cm}\sum\limits^{4}_{\mu\hspace{0.02cm} = \hspace{0.02cm}1}p^{(k)}_{\mu}x^{(k - 1)}_{\mu} +
\frac{1}{2}\,\sum\limits^{2}_{l=1}\,
[\hspace{0.04cm}\bar{\xi}^{\,\prime(k)}_{l}\!,
\hspace{0.02cm}\xi^{\phantom{\prime}(k - 1)}_{l}\hspace{0.03cm}]
\biggr\}.
\]
We recall that in (\ref{eq:3g}) the Dirac brackets designate
\[
\langle\hspace{0.02cm}(k)^{\prime}_{p}\hspace{0.03cm}|
\equiv
\langle\hspace{0.02cm}p^{(k)}, \bar{\xi}^{\hspace{0.02cm}\prime(k)}\hspace{0.02cm}|\hspace{0.02cm},
\qquad
|\hspace{0.03cm}(k - 1)_{x}\hspace{0.02cm}\rangle
\equiv
|\,x^{(k- 1)},\xi^{(k - 1)}\hspace{0.02cm}\rangle.
\]
In the given paper we restrict our consideration to an analysis of the matrix element of term linear with respect to the covariant derivative, i.e. of the term $[\hspace{0.02cm}\chi, \hat{\cal L}(z, D)\hspace{0.02cm}]$ in (\ref{eq:3g}). The calculation of the matrix elements for more complicated contributions $\hat{H}(z)$ and $[\hspace{0.02cm}\chi, \hat{\cal L}(z, D)\hspace{0.02cm}]^{\hspace{0.02cm}2}$ will be presented in Part II \cite{part_II} after the development of all required mathematical technique.\\
\indent At the end of this section we write out in an expanded form the expression for the term $\hat{\eta}_{\mu}(z) \hat{D}_{\mu}$, which is included into the definition of the operator $\hat{\cal L}(z, \hat{D})$, Eq.\,(\ref{eq:3q}). Taking into account (\ref{eq:2u}), we get
\begin{equation}
\hat{\eta}_{\mu}(z)\hspace{0.03cm}\hat{D}_{\mu}
=
\biggl(1 + \frac{1}{2}\,z\!\hspace{0.03cm}\biggr)\hat{\beta}_{\mu}\hat{D}_{\mu}
+
z\hspace{0.03cm}\biggl(\frac{i\sqrt{3}}{2}\hspace{0.02cm}\biggr)
\hspace{0.03cm}[\hspace{0.03cm}\hat{\omega}, \hat{\beta}_{\mu}\hspace{0.02cm}]
\hspace{0.03cm}\hat{D}_{\mu}
=
\label{eq:3h}
\end{equation}
\[
=
\frac{1}{2}\,\sum\limits^{2}_{n\hspace{0.02cm}=\hspace{0.02cm}1}
\,\biggl\{\!\hspace{0.01cm}\biggl(1 + \frac{1}{2}\,z\!\hspace{0.03cm}\biggr)
\hspace{0.02cm}\bigl(\hat{D}^{\phantom{-}}_{\bar{n}}\hspace{0.03cm}a^{-}_{n}
+
\hat{D}^{\phantom{+}}_{n}\hspace{0.03cm}a^{+}_{n}\bigr)
+
z\hspace{0.03cm}\biggl(\frac{i\sqrt{3}}{2}\hspace{0.02cm}\biggr)\hspace{0.02cm}
\bigl(\hat{D}^{\phantom{-}}_{\bar{n}}\hspace{0.03cm}[\hspace{0.03cm}\hat{\omega}, a^{-}_{n}\hspace{0.02cm}]
+
\hat{D}^{\phantom{+}}_{n}\hspace{0.03cm}[\hspace{0.03cm}\hat{\omega}, a^{+}_{n}\hspace{0.02cm}]\hspace{0.02cm}\bigr)\!\hspace{0.02cm}\biggr\},
\]
where
\[
\hat{D}_{\bar{n}} = -i\hspace{0.03cm}\bigl(\hat{P}_{\bar{n}} - e\hspace{0.02cm}A_{\bar{n}}(\hat{x})\bigr),
\quad
\hat{D}_{n} = -i\hspace{0.03cm}\bigl(\hat{P}_{n} - e\hspace{0.02cm}A_{n}(\hat{x})\bigr).
\]
In (\ref{eq:3h}) we have turned to the creation and annihilation operators in accordance with (\ref{eq:3r}).


\section{\bf Matrix element  \texorpdfstring{$\langle\hspace{0.02cm}(k)^{\prime}_{p}\hspace{0.02cm}|\hspace{0.02cm}
[\hspace{0.03cm}\chi,\hat{\cal L}(z,\hat{D})\hspace{0.01cm}]\hspace{0.01cm}|\hspace{0.02cm}
(k - 1)_{x}\hspace{0.02cm}\rangle$}{a0a2}}
\setcounter{equation}{0}
\label{section_4}

In this section we give a detail form for the matrix element of term linear in the operator $\hat{\cal L}(z,\hat{D})$ in the general expression (\ref{eq:3g}). Since the variable $\chi$ is a para-Grassmann number, then by virtue of relation (\ref{ap:B5}) and definition of the parafermion coherent states (\ref{eq:3i}), it can be factored out from the Dirac brackets $\langle(k)_{p}^{\prime}\vert$ and $\vert\,(k-1)_x\rangle$:
\begin{equation}
\langle\hspace{0.02cm}(k)^{\prime}_{p}\hspace{0.02cm}|\hspace{0.02cm}
[\hspace{0.03cm}\chi\hspace{0.02cm},
\hat{\cal L}(z,\hat{D})\hspace{0.01cm}]\hspace{0.01cm}|\hspace{0.02cm}
(k - 1)_{x}\hspace{0.02cm}\rangle
=
[\hspace{0.03cm}\chi\hspace{0.02cm},
\langle\hspace{0.02cm}(k)^{\prime}_{p}\hspace{0.02cm}|\hspace{0.02cm}
\hat{\cal L}(z,\hat{D})\hspace{0.01cm}|\hspace{0.02cm}(k - 1)_{x}\hspace{0.02cm}\rangle
\hspace{0.01cm}]
=
\label{eq:4q}
\end{equation}
\[
=
\frac{\!i}{\,\varepsilon^{1/3}(z)}\,[\hspace{0.03cm}\chi\hspace{0.02cm},
\langle\hspace{0.02cm}(k)^{\prime}_{p}\hspace{0.02cm}|\hspace{0.03cm}
\hat{A}\hspace{0.03cm}\hat{\eta}_{\mu}(z)\hspace{0.02cm} \hat{D}_{\mu}\hspace{0.01cm}|\hspace{0.02cm}(k - 1)_{x}\hspace{0.02cm}\rangle
\hspace{0.01cm}]
+
m\hspace{0.03cm}[\hspace{0.03cm}\chi\hspace{0.02cm},
\langle\hspace{0.02cm}(k)^{\prime}_{p}\hspace{0.02cm}|\hspace{0.03cm}
\hat{A}\hspace{0.01cm}|\hspace{0.03cm}(k - 1)_{x}\hspace{0.02cm}\rangle
\hspace{0.01cm}].
\]
By using the representation (\ref{eq:3h}) for the first term in the last line, we have
\begin{equation}
\langle\hspace{0.02cm}(k)^{\prime}_{p}\hspace{0.02cm}|\hspace{0.03cm}
\hat{A}\hspace{0.03cm}\hat{\eta}_{\mu}(z)\hspace{0.02cm} \hat{D}_{\mu}\hspace{0.01cm}|\hspace{0.02cm}(k - 1)_{x}\hspace{0.02cm}\rangle
=
\label{eq:4w}
\end{equation}
\[
\begin{split}
=
\frac{1}{2}\,\sum\limits^{2}_{n\hspace{0.02cm}=\hspace{0.02cm}1}
\,\biggl\{\!\hspace{0.01cm}&\biggl(1 + \frac{1}{2}\,z\!\hspace{0.03cm}\biggr)
\langle\hspace{0.02cm}(k)^{\prime}_{p}\hspace{0.02cm}|\hspace{0.03cm}
\hat{A}\hspace{0.02cm}\bigl(\hat{D}^{\phantom{-}\!}_{\bar{n}}\hspace{0.03cm}a^{-}_{n}  
+ 
\hat{D}^{\phantom{+}\!}_{n}\hspace{0.03cm}a^{+}_{n}\bigr)|\hspace{0.03cm}
(k - 1)_{x}\hspace{0.02cm}\rangle
\,+ \\[1ex]
+\,
&z\hspace{0.03cm}\biggl(\frac{i\sqrt{3}}{2}\hspace{0.02cm}\biggr)\hspace{0.02cm}
\langle\hspace{0.02cm}(k)^{\prime}_{p}\hspace{0.02cm}|\hspace{0.04cm}
\hat{A}\bigl(\hat{D}^{\phantom{-}\!}_{\bar{n}}\hspace{0.03cm}[\hspace{0.03cm}\hat{\omega}, a^{-}_{n}\hspace{0.02cm}]
+
\hat{D}^{\phantom{+}\!}_{n}\hspace{0.03cm}[\hspace{0.03cm}\hat{\omega}, a^{+}_{n}\hspace{0.02cm}]\hspace{0.02cm}\bigr)|\hspace{0.04cm}
(k - 1)_{x}\hspace{0.02cm}\rangle\!\hspace{0.03cm}\biggr\}
=
\end{split}
\vspace{-0.5cm}
\]
\begin{align}
=\! -\frac{i}{2}\Biggl[\hspace{0.02cm}&\sum\limits^{2}_{n\hspace{0.02cm}=\hspace{0.02cm}1}
\hspace{0.02cm}
\biggl\{\!\hspace{0.01cm}\biggl(1 + \frac{1}{2}\,z\!\hspace{0.03cm}\biggr)
\langle\hspace{0.02cm}\bar{\xi}^{\,\prime(k)}\hspace{0.02cm}|\,
\hat{A}\hspace{0.03cm}a^{-}_{n}|\,\xi^{(k - 1)}\hspace{0.02cm}\rangle
\!+\!
z\hspace{0.03cm}\biggl(\frac{i\sqrt{3}}{2}\hspace{0.02cm}\biggr)
\langle\hspace{0.02cm}\bar{\xi}^{\,\prime(k)}\hspace{0.02cm}|\hspace{0.02cm}
\hat{A}\hspace{0.03cm}[\hspace{0.03cm}\hat{\omega}, a^{-}_{n}\hspace{0.02cm}]
\hspace{0.02cm}|\,\xi^{(k - 1)}\hspace{0.02cm}\rangle\!\biggr\}
\bigl(p^{(k)}_{\bar{n}}\!- e\hspace{0.02cm}A_{\bar{n}}(x^{(k-1)})\bigr)
\notag \\[1ex]
+
&\sum\limits^{2}_{n\hspace{0.02cm}=\hspace{0.02cm}1}\hspace{0.02cm}
\biggl\{\!\hspace{0.01cm}\biggl(1 + \frac{1}{2}\,z\!\hspace{0.03cm}\biggr)
\langle\hspace{0.02cm}\bar{\xi}^{\,\prime(k)}\hspace{0.02cm}|
\hat{A}\hspace{0.03cm}a^{+}_{n}|\,\xi^{(k - 1)}\hspace{0.02cm}\rangle
\!+\!
z\hspace{0.03cm}\biggl(\frac{i\sqrt{3}}{2}\hspace{0.02cm}\biggr)
\langle\hspace{0.02cm}\bar{\xi}^{\,\prime(k)}\hspace{0.02cm}|\hspace{0.02cm}
\hat{A}\hspace{0.03cm}[\hspace{0.03cm}\hat{\omega}, a^{+}_{n}\hspace{0.02cm}]
\hspace{0.02cm}|\,\xi^{(k - 1)}\hspace{0.02cm}\rangle\!\biggr\}
\bigl(p^{(k)}_{n}\!- e\hspace{0.02cm}A_{n}(x^{(k-1)})\bigr)
\Biggr] \notag
\end{align}
\[
\times\,\langle\hspace{0.02cm}p^{(k)}|\,x^{(k-1)\hspace{0.02cm}}\rangle.
\]
Matrix element in the mass term on the right-hand side of (\ref{eq:4q}) by virtue of the expansion (\ref{eq:2p}) has the following form:
\begin{equation}
\langle\hspace{0.02cm}(k)^{\prime}_{p}\hspace{0.02cm}|\hspace{0.03cm}
\hat{A}\hspace{0.01cm}|\hspace{0.02cm}(k - 1)_{x}\hspace{0.02cm}\rangle
=
\label{eq:4e}
\end{equation}
\[
=
\left(\alpha\hspace{0.03cm}\langle\hspace{0.02cm}\bar{\xi}^{\,\prime(k)}\hspace{0.02cm}|
\,\xi^{(k - 1)}\hspace{0.02cm}\rangle
+
\beta\hspace{0.03cm}\langle\hspace{0.02cm}\bar{\xi}^{\,\prime(k)}\hspace{0.02cm}|
\;\hat{\omega}\hspace{0.03cm}|\,\xi^{(k - 1)}\hspace{0.02cm}\rangle
+
\gamma\hspace{0.03cm}\langle\hspace{0.02cm}\bar{\xi}^{\,\prime(k)}\hspace{0.02cm}|
\;\hat{\omega}^{\hspace{0.02cm}2}\hspace{0.03cm}|\,\xi^{(k - 1)}\hspace{0.02cm}\rangle
\right)
\langle\hspace{0.02cm}p^{(k)}|\,x^{(k-1)\hspace{0.02cm}}\rangle.
\]
Thus in analysis of the expression (\ref{eq:4q}) we face with the necessity of calculating matrix elements for the operators $\hat{\omega},\, \hat{\omega}^{\hspace{0.03cm}2},\, \hat{A}\hspace{0.03cm}a_{n}^{\pm}$ and $\hat{A}\hspace{0.03cm}[\hspace{0.03cm}\hat{\omega}, a_n^{\pm}\hspace{0.02cm}]$ in the basis of parafermion coherent states. We carry this out in several stages, the first of which is to define a connection between the operator $\hat{\omega}$, which within the framework of the DKP theory is given by expression (\ref{eq:2i}) and the operator $a_{0}$ arising in the scheme of quantization based on the Lie algebra of the orthogonal group $SO(2M + 2)$, Eq.\,(\ref{ap:A3}).


\section{Operator \texorpdfstring{$a_{0}$}{a0a2}}
\setcounter{equation}{0}
\label{section_5}

For convenience of further references we write out all independent state vectors spanned by the operators 
$a_{k}^{+}$. These states are
\begin{equation}
\begin{array}{llll}
&\mbox{null-particle state}\!: &|\hspace{0.03cm}0\rangle, \\[1ex]
&\mbox{one-particle states}\!:
&|\hspace{0.03cm}1\rangle\equiv a^{+}_{1} |\hspace{0.03cm}0\rangle,
&|\hspace{0.03cm}2\rangle\equiv a^{+}_{2} |\hspace{0.03cm}0\rangle,
\\[1ex]
&\mbox{two-particle states}\!:
&|\hspace{0.03cm}11\rangle\equiv (a^{+}_{1})^{2} |\hspace{0.03cm}0\rangle,
&|\hspace{0.03cm}22\rangle\equiv (a^{+}_{2})^{2} |\hspace{0.03cm}0\rangle,\\[1ex]
&\phantom{\mbox{two-particle states}\!:}
&|\hspace{0.03cm}12\rangle\equiv a^{+}_{1}a^{+}_{2} |\hspace{0.03cm}0\rangle,
&|\hspace{0.03cm}21\rangle\equiv a^{+}_{2}a^{+}_{1} |\hspace{0.03cm}0\rangle,
\\[1ex]
&\mbox{three-particle states}\!:
&|\hspace{0.03cm}112\rangle\equiv (a^{+}_{1})^{2}a^{+}_{2} |\hspace{0.03cm}0\rangle,
\hspace{-1cm}
&|\hspace{0.03cm}221\rangle\equiv (a^{+}_{2})^{2} a^{+}_{1} |\hspace{0.03cm}0\rangle,
\\[1ex]
&\mbox{four-particle state}\!:
&|\hspace{0.03cm}1122\rangle\equiv (a^{+}_{1})^{2} (a^{+}_{2})^{2} |\hspace{0.03cm}0\rangle.
\end{array}
\label{eq:5q}
\end{equation}
\indent Let us define the rules of an action of the operator $a_0$ on the state vectors (\ref{eq:5q}). For definiteness we fix the positive sign in formula (\ref{ap:A5}), i.e. we set
\begin{equation}
a_{0}\hspace{0.01cm}|\hspace{0.03cm}0\rangle = |\hspace{0.03cm}0\rangle.
\label{eq:5w}
\end{equation}
Hereinafter, for convenience of further construction we redefine the operator $a_0$: $a_{0}\rightarrow 2\hspace{0.02cm}a_{0}$.
Then from general relations (\ref{ap:A6}) with allowance for algebra (\ref{eq:3t})\,--\,(\ref{eq:3u}) it follows that
\begin{align}
&a_{0}\hspace{0.01cm}|\hspace{0.03cm}1\rangle = a_{0}\hspace{0.01cm}|\hspace{0.03cm}2\rangle = 0,
\label{eq:5e} \\[1ex]
&a_{0}\hspace{0.01cm}|\hspace{0.03cm}11\rangle = - |\hspace{0.03cm}11\rangle, \quad
a_{0}\hspace{0.01cm}|\hspace{0.03cm}22\rangle = - |\hspace{0.03cm}22\rangle,
\label{eq:5r} \\[1ex]
&a_{0}\hspace{0.01cm}|\hspace{0.03cm}12\rangle = - |\hspace{0.03cm}21\rangle, \quad
a_{0}\hspace{0.01cm}|\hspace{0.03cm}21\rangle = - |\hspace{0.03cm}12\rangle,
\label{eq:5t} \\[1ex]
&a_{0}\hspace{0.01cm}|\hspace{0.03cm}112\rangle = a_{0}\hspace{0.01cm}|\hspace{0.03cm}221\rangle = 0,
\label{eq:5y} \\[1ex]
&a_{0}\hspace{0.01cm}|\hspace{0.03cm}1122\rangle = |\hspace{0.03cm}1122\rangle. 
\label{eq:5u}
\end{align}
The operator $a_{0}$ turns into zero states with an odd number of parafermions. The signs on the right-hand side (\ref{eq:5r}), (\ref{eq:5t}) and (\ref{eq:5u}) are connected with a choice of the sign in (\ref{eq:5w}). The relation (\ref{eq:5t}) is of special interest. Two different states $|\hspace{0.03cm}12\rangle$ and $|\hspace{0.03cm}21\rangle$ are orthogonal to each other and contain the same number of parafermions of sorts 1 and 2, i.e. the two-particle system has a two-fold degeneracy. The operator $a_{0}$ correct to a sign changes one state to another.\\
\indent Now we consider the question of obtaining an explicit form of the operator $a_{0}$ in terms of the generators $L_{kl},\,M_{kl}$ and $N_{kl}$ of the group $SO(2M)$ as they were defined in the paper by Kamefuchi and Takahashi \cite{kamefuchi_1962}. In the special case $M = 2$ we have the following components of these generators different from zero:
\begin{equation}
\begin{array}{llll}
&L^{\phantom{+\!\!}}_{12} = \displaystyle\frac{1}{2}\;[\hspace{0.03cm}a^{+}_{1}, a^{+}_{2}\hspace{0.02cm}],
\quad\quad
&M^{\phantom{+\!\!}}_{12} = \displaystyle\frac{1}{2}\;[\hspace{0.03cm}a^{-}_{1}, a^{-}_{2}\hspace{0.02cm}], \\[2ex]
&N^{\phantom{+\!\!}}_{12} = \displaystyle\frac{1}{2}\;[\hspace{0.03cm}a^{+}_{1}, a^{-}_{2}\hspace{0.02cm}],
\quad\quad
&N^{\phantom{+\!\!}}_{21} = \displaystyle\frac{1}{2}\;[\hspace{0.03cm}a^{+}_{2}, a^{-}_{1}\hspace{0.02cm}], \\[2ex]
&N^{\phantom{+\!\!}}_{1\,} = \displaystyle\frac{1}{2}\;[\hspace{0.03cm}a^{+}_{1}, a^{-}_{1}\hspace{0.02cm}],
\quad\quad
&N^{\phantom{+\!\!}}_{\,2\,} = \displaystyle\frac{1}{2}\;[\hspace{0.03cm}a^{+}_{2}, a^{-}_{2}\hspace{0.02cm}].
\end{array}
\label{eq:5i}
\end{equation}
It is shown by us\footnote{\hspace{0.03cm}All technical details of calculations here and in the subsequent text of the present paper will be given in our special work \cite{markov_2020}.} that the operator $a_0$ as a function of the generators (\ref{eq:5i}), correctly reproducing the relations (\ref{eq:5w})\,--\,(\ref{eq:5u}) has of the following structure:
\begin{equation}
a_{0} = -\frac{1}{4}\,\bigl(\{\hspace{0.02cm}L_{12}, M_{12}\hspace{0.02cm}\}
+
\{\hspace{0.02cm}N_{12}, N_{21}\hspace{0.02cm}\}
-
\{\hspace{0.02cm}N_{1}, N_{2}\hspace{0.02cm}\}\bigr).
\label{eq:5o}
\end{equation}
In deriving this expression the general commutation relations for the generators $L_{kl},\, M_{kl}$ and $N_{kl}$ from \cite{kamefuchi_1962}  for $M = 2$ and the commutation relations with the operators $a_{k}^{\pm}$ were used:
\begin{equation}
\begin{array}{llll}
&[\hspace{0.03cm}a^{-}_{k}, L^{\phantom{-}}_{12}\hspace{0.02cm}]
= \delta^{\phantom{+}}_{k1}\hspace{0.03cm}a^{+}_{2}
- \delta^{\phantom{+}}_{k2}\hspace{0.03cm}a^{+}_{1},
\quad
&[\hspace{0.03cm}a^{-}_{k}, M^{\phantom{+\!\!}}_{12}\hspace{0.02cm}] = 0, \\[1ex]
&[\hspace{0.03cm}a^{+}_{k}, M^{\phantom{+}}_{12}\hspace{0.02cm}]
= \delta^{\phantom{-}}_{k1}\hspace{0.03cm}a^{-}_{2}
- \delta^{\phantom{-}}_{k2}\hspace{0.03cm}a^{-}_{1},
\quad
&[\hspace{0.03cm}a^{+}_{k}, L^{\phantom{+\!\!}}_{12}\hspace{0.02cm}] = 0,\\[1ex]
&[\hspace{0.03cm}a^{-}_{k}, N^{\phantom{+\!\!}}_{12}\hspace{0.02cm}] =
\delta_{k1}\hspace{0.03cm}a^{-}_{2},
\quad
&[\hspace{0.03cm}a^{+}_{k}, N^{\phantom{+}}_{12}\hspace{0.02cm}] = -\hspace{0.02cm}\delta^{\phantom{+}}_{k2}\hspace{0.03cm}a^{+}_{1}, \\[1ex]
&[\hspace{0.03cm}a^{-}_{k}, N^{\phantom{+\!\!}}_{21}\hspace{0.02cm}] =
\delta_{k2}\hspace{0.03cm}a^{-}_{1},
\quad
&[\hspace{0.03cm}a^{+}_{k}, N^{\phantom{+}}_{21}\hspace{0.02cm}] = -\hspace{0.02cm}\delta^{\phantom{+}}_{k1}\hspace{0.03cm}a^{+}_{2}, \\[1ex]
&[\hspace{0.03cm}a^{-}_{l}, N^{\phantom{+\!\!}}_{k}\hspace{0.02cm}] =
\delta_{kl}\hspace{0.03cm}a^{-}_{l},
\quad
&[\hspace{0.03cm}a^{+}_{l}, N^{\phantom{+}}_{k}\!] = -\hspace{0.02cm}\delta^{\phantom{+}}_{kl}\hspace{0.03cm}a^{+}_{l}.
\end{array}
\label{eq:5p}
\end{equation}
\indent Let us define a connection between the operator $a_{0}$ and the matrix $\omega$, Eq.\,(\ref{eq:2i}). In particular, for $M = 2$ within the framework of the operator formalism, we have
\[
\hat{\omega} \,= \frac{1}{4}\; \epsilon_{\mu\nu\lambda\sigma}\hspace{0.02cm}
\hat{\beta}_{\mu}\hspace{0.02cm}\hat{\beta}_{\nu}\hspace{0.02cm}
\hat{\beta}_{\lambda}\hspace{0.02cm}\hat{\beta}_{\sigma}
\]
or in an equivalent form:
\[
\hat{\omega} \,= \biggl(\frac{1}{4}\biggr)^{\!\!2} \epsilon_{\mu\nu\lambda\sigma}\hspace{0.02cm}
[\hspace{0.02cm}\hat{\beta}_{\mu}, \hat{\beta}_{\nu}\hspace{0.02cm}]\hspace{0.02cm}
[\hspace{0.02cm}\hat{\beta}_{\lambda}, \hat{\beta}_{\sigma}\hspace{0.02cm}]
=
\]
\begin{equation}
=\frac{1}{4}\,\Bigl(\bigl\{
[\hspace{0.02cm}\hat{\beta}_{1}, \hat{\beta}_{2}\hspace{0.02cm}],
[\hspace{0.02cm}\hat{\beta}_{3}, \hat{\beta}_{4}\hspace{0.02cm}]\bigr\}
+
[\hspace{0.02cm}\hat{\beta}_{1}, \hat{\beta}_{4}\hspace{0.02cm}],
[\hspace{0.02cm}\hat{\beta}_{2}, \hat{\beta}_{3}\hspace{0.02cm}]\bigr\}
-
[\hspace{0.02cm}\hat{\beta}_{1}, \hat{\beta}_{3}\hspace{0.02cm}],
[\hspace{0.02cm}\hat{\beta}_{2}, \hat{\beta}_{4}\hspace{0.02cm}]\bigr\}\!
\Bigr).
\label{eq:5a}
\end{equation}
We rewrite the expression in the last line in terms of the creation and annihilation operators by using the connection (\ref{eq:3r}) and the definitions (\ref{eq:5i}). It is easy to show that the following relations hold:
\[
\begin{split}
&[\hspace{0.02cm}\hat{\beta}_{1}, \hat{\beta}_{2}\hspace{0.02cm}] = i\hspace{0.015cm}N_{1},
\quad
[\hspace{0.02cm}\hat{\beta}_{3}, \hat{\beta}_{4}\hspace{0.02cm}] =
i\hspace{0.015cm}N_{2}, \\[1ex]
&[\hspace{0.02cm}\hat{\beta}_{1}, \hat{\beta}_{4}\hspace{0.02cm}] =
\frac{1}{2\hspace{0.02cm}i}\,\bigl[\hspace{0.02cm}(L_{12} - M_{12}) - (N_{12} + N_{21})\bigr], \\[1ex]
&[\hspace{0.02cm}\hat{\beta}_{2}, \hat{\beta}_{3}\hspace{0.02cm}] =
\frac{1}{2\hspace{0.02cm}i}\,\bigl[\hspace{0.02cm}(L_{12} - M_{12}) + (N_{12} + N_{21})\bigr], \\[1ex]
&[\hspace{0.02cm}\hat{\beta}_{1}, \hat{\beta}_{3}\hspace{0.02cm}] =
\frac{1}{2}\,\bigl[\hspace{0.02cm}(L_{12} + M_{12}) + (N_{12} - N_{21})\bigr], \\[1ex]
&[\hspace{0.02cm}\hat{\beta}_{2}, \hat{\beta}_{4}\hspace{0.02cm}] =
-\frac{1}{2}\,\bigl[\hspace{0.02cm}(L_{12} + M_{12}) - (N_{12} - N_{21})\bigr].
\end{split}
\]
Substituting these expressions into (\ref{eq:5a}), we obtain an explicit form of the operator $\hat{\omega}$ in terms of generators of the orthogonal group $SO(4)$
\begin{equation}
\hat{\omega} \,= \frac{1}{4}\,\bigl(\{\hspace{0.02cm}L_{12}, M_{12}\hspace{0.02cm}\}
+
\{\hspace{0.02cm}N_{12}, N_{21}\hspace{0.02cm}\}
-
\{\hspace{0.02cm}N_{1}, N_{2}\hspace{0.02cm}\}\bigr).
\label{eq:5s}
\end{equation}
Comparing (\ref{eq:5s}) with (\ref{eq:5o}), we get the desired relation between the operators 
$\hat{\omega}$ and $a_{0}$
\[
\hat{\omega} \,= - \hspace{0.03cm}a_{0}.
\]
The minus sign on the right-hand side is caused by  the choice of the sign in (\ref{eq:5w}). From the previous relation and from the property (\ref{eq:2o}) it immediately follows that
\[
a^{3}_{0} = a^{\phantom{3}}_{0}.
\]
\indent Given an explicit form of the operator $a_0$, Eq.\,(\ref{eq:5o}), we can define its matrix element in the basis of the para-Fermi coherent states. The desired matrix element of the operator $a_{0}$ has the following form (details of calculations see in \cite{markov_2020}):
\begin{equation}
\begin{split}
\langle\hspace{0.02cm}\bar{\xi}^{\,\prime}\hspace{0.02cm}|\,
a_{0}|\,\xi\hspace{0.02cm}\rangle
=
-\frac{1}{2}\,\biggl\{\!
&\biggl(\displaystyle\frac{1}{2}\;[\hspace{0.03cm}\bar{\xi}^{\,\prime}_{1}, \bar{\xi}^{\,\prime}_{2}\hspace{0.03cm}]\biggr)\!
\biggl(\,\displaystyle\frac{1}{2}\;[\hspace{0.03cm}\xi^{\phantom{\prime}}_{1}, \xi^{\phantom{\prime}}_{2}\hspace{0.03cm}]\biggr)
+
\biggl(\displaystyle\frac{1}{2}\;[\hspace{0.03cm}\bar{\xi}^{\,\prime}_{1}, \xi^{\phantom{\prime}}_{2}\hspace{0.03cm}]\biggr)\!
\biggl(\displaystyle\frac{1}{2}\;[\hspace{0.03cm}\bar{\xi}^{\,\prime}_{2}, \xi^{\phantom{\prime}}_{1}\hspace{0.03cm}]\biggr) - \\[1ex]
-\, &\biggl(\displaystyle\frac{1}{2}\;[\hspace{0.03cm}\bar{\xi}^{\,\prime}_{1}, \xi^{\phantom{\prime}}_{1}\hspace{0.03cm}]\biggr)\!
\biggl(\displaystyle\frac{1}{2}\;[\hspace{0.03cm}\bar{\xi}^{\,\prime}_{2}, \xi^{\phantom{\prime}}_{2}\hspace{0.03cm}]\biggr)
+
2\biggl(\displaystyle\frac{1}{2}\;[\hspace{0.03cm}\bar{\xi}^{\,\prime}_{1}, \xi^{\phantom{\prime}}_{1}\hspace{0.03cm}]
+
\displaystyle\frac{1}{2}\;[\hspace{0.03cm}\bar{\xi}^{\,\prime}_{2}, \xi^{\phantom{\prime}}_{2}\hspace{0.03cm}] - 1\biggr)\!
\biggr\}\hspace{0.03cm}
\langle\hspace{0.02cm}\bar{\xi}^{\,\prime}\hspace{0.02cm}|\,\xi\hspace{0.02cm}\rangle.
\end{split}
\label{eq:5f}
\end{equation}


\section{Matrix element of the operator \texorpdfstring{$a^{2}_{0}$}{a0a2}}
\setcounter{equation}{0}
\label{section_6}

The explicit form of the operator $a_0^2$ is given in Appendix \ref{appendix_A}, Eq.\,(\ref{ap:A7}). If one introduce the para-Fermi number operator of the $k$ state (for parastatistics $p = 2$)
\[
n^{\phantom{+\!\!\!}}_{k\,} = \displaystyle\frac{1}{2}\;[\hspace{0.03cm}a^{+}_{k}, a^{-}_{k}\hspace{0.02cm}] + 1
= N^{\phantom{+\!\!}}_{k} + 1,
\]
then the expression (\ref{ap:A7}) can be presented in the following form:
\begin{equation}
a^{2}_{0} = 1 - \bigl\{(n^{\phantom{2\!\!}}_{1} - 1)^{2} + (n^{\phantom{2\!\!}}_{2} - 1)^{2}\bigr\}
+
2\hspace{0.03cm} (n^{\phantom{2\!\!}}_{1} - 1)^{2}(n^{\phantom{2\!\!}}_{2} - 1)^{2}.
\label{eq:6w}
\end{equation}
\indent Let us determine action of the operator $a_{0}^{2}$ on the coherent state (\ref{eq:3i}). For this purpose, we find a rule of action of the para-Fermi number operator $n_{k}$ on $|\,\xi\hspace{0.02cm}\rangle$. By using the operator identity
\[
{\rm e}^{X}Y{\rm e}^{-X} = Y + [X,Y\hspace{0.02cm}] + \frac{\!1}{2!}\hspace{0.03cm}[X,[X,Y\hspace{0.02cm}]\hspace{0.02cm}]
+ \frac{\!1}{3!}\hspace{0.03cm}[X,[X,[X,Y\hspace{0.02cm}]\hspace{0.02cm}]\hspace{0.02cm}] + \ldots
\]
and commutation relations (\ref{ap:B3}) and (\ref{ap:B4})  it is easy to show that the following equality takes place
\begin{equation}
n^{\phantom{+\!\!}}_{k}\hspace{0.02cm}|\,\xi\hspace{0.02cm}\rangle =
\biggl(\displaystyle\frac{1}{2}\;[\hspace{0.03cm}a^{+}_{k}, \xi^{\phantom{+}\!}_{k}\hspace{0.02cm}]\biggr) |\,\xi\hspace{0.02cm}\rangle.
\label{eq:6r}
\end{equation}
Recall that there is no summation over repeated Latin indices. Similar calculation for $n_{k}^{2}$ gives
\begin{equation}
n^{2}_{k}\hspace{0.02cm}|\,\xi\hspace{0.02cm}\rangle =
\biggl\{\displaystyle\frac{1}{2}\;[\hspace{0.03cm}a^{+}_{k}, \xi^{\phantom{+}\!}_{k}\hspace{0.02cm}]
+
\biggl(\displaystyle\frac{1}{2}\;[\hspace{0.03cm}a^{+}_{k}, \xi^{\phantom{+}\!}_{k}\hspace{0.02cm}]\biggr)^{\!\!2\,}
\biggr\} |\,\xi\hspace{0.02cm}\rangle.
\label{eq:6t}
\end{equation}
In view of the definition (\ref{eq:6w}), it follows from (\ref{eq:6r}) and (\ref{eq:6t}) that
\[
a^{2}_{0}|\,\xi\hspace{0.02cm}\rangle =
\biggl\{1 - \sum\limits^{2}_{k = 1}
\biggl[\biggl(\displaystyle\frac{1}{2}\;[\hspace{0.03cm}a^{+}_{k}, \xi^{\phantom{+}\!}_{k}\hspace{0.02cm}]\biggr)^{\!\!2\,} -
 \displaystyle\frac{1}{2}\;[\hspace{0.03cm}a^{+}_{k}, \xi^{\phantom{+}\!}_{k}\hspace{0.02cm}]
+ 1\biggr]
+
2\prod\limits^{2}_{k = 1}
\biggl[\biggl(\displaystyle\frac{1}{2}\;[\hspace{0.03cm}a^{+}_{k}, \xi^{\phantom{+}\!}_{k}\hspace{0.02cm}]\biggr)^{\!\!2\,} -
 \displaystyle\frac{1}{2}\;[\hspace{0.03cm}a^{+}_{k}, \xi^{\phantom{+}\!}_{k}\hspace{0.02cm}]
+ 1\biggr]
\biggr\} |\,\xi\hspace{0.02cm}\rangle,
\]
and thus the required matrix element has the form
\begin{equation}
\begin{split}
\langle\hspace{0.02cm}\bar{\xi}^{\,\prime}\hspace{0.02cm}|\,
a^{2}_{0}\hspace{0.03cm}|\,\xi\hspace{0.02cm}\rangle
=
\biggl\{1 - \sum\limits^{2}_{k = 1}\hspace{0.02cm}
&\biggl[\biggl(\displaystyle\frac{1}{2}\;[\hspace{0.03cm}\bar{\xi}^{\,\prime}_{k}, \xi^{\phantom{\prime}}_{k}\hspace{0.02cm}]\biggr)^{\!\!2\,} -
\displaystyle\frac{1}{2}\;[\hspace{0.03cm}\bar{\xi}^{\,\prime}_{k}, \xi^{\phantom{\prime}}_{k}\hspace{0.02cm}]
+ 1\hspace{0.03cm}\biggr]\, + \\[1ex]
+\, 2\hspace{0.02cm}\prod\limits^{2}_{k = 1}\hspace{0.02cm}
&\biggl[\biggl(\displaystyle\frac{1}{2}\;[\hspace{0.03cm}\bar{\xi}^{\,\prime}_{k}, \xi^{\phantom{\prime}}_{k}\hspace{0.02cm}]\biggr)^{\!\!2\,} -
\displaystyle\frac{1}{2}\;[\hspace{0.03cm}\bar{\xi}^{\,\prime}_{k}, \xi^{\phantom{\prime}}_{k}\hspace{0.02cm}]
+ 1\hspace{0.03cm}\biggr]\biggr\}
\langle\hspace{0.02cm}\bar{\xi}^{\,\prime}\hspace{0.02cm}|\,\xi\hspace{0.02cm}\rangle.
\end{split}
\label{eq:6y}
\end{equation}
This matrix element along with the matrix element for the operator $a_{0}$, Eq.\,(\ref{eq:5f}), enables us to fully define the expression for matrix element of operator $\hat{A}$ as it was defined by Eq.\,(\ref{eq:4e}) (with the replacement $\hat{\omega}\rightarrow -\hspace{0.02cm}a_{0}$). In particular, from here we have immediately a consequence. By virtue of the fact that the expressions (\ref{eq:5f}) and (\ref{eq:6y}) are defined only by the commutators of para-Grassmann numbers, the first relation in \eqref{ap:B1} leads to that the last term in (\ref{eq:4q}) vanishes, i.e.
\[
[\hspace{0.03cm}\chi, \langle\hspace{0.02cm}\bar{\xi}^{\,\prime(k)}\hspace{0.02cm}|\hspace{0.03cm}
\hat{A}\hspace{0.01cm}|\,\xi^{(k -1)}\hspace{0.02cm}\rangle\hspace{0.02cm}] = 0.
\]
Finally, it is possible to show that the matrix element (\ref{eq:6y}) can be presented in very simple form \cite{markov_2020}
\[
\langle\hspace{0.02cm}\bar{\xi}^{\,\prime}\hspace{0.02cm}|\,
a^{2}_{0}\,|\,\xi\hspace{0.02cm}\rangle
=
\cosh\hspace{0.03cm}\Bigl(\frac{1}{2}\sum_{l}\,
[\hspace{0.03cm}\bar{\xi}^{\,\prime}_{l}\hspace{0.02cm},\xi^{\phantom{\prime}\!}_{l}
\hspace{0.02cm}\hspace{0.02cm}]\Bigr).
\]


\section{Matrix elements of the commutators \texorpdfstring{$[\hspace{0.03cm}a^{ }_{0}, a^{\pm}_{n}\hspace{0.02cm}]$ and $[\hspace{0.03cm}a^{2}_{0}, a^{\pm}_{n}\hspace{0.02cm}]$}{a0a2}}
\setcounter{equation}{0}
\label{section_7}

Let us return to matrix element (\ref{eq:4w}). The first term in braces on the right-hand side has the form
\begin{equation}
\langle\hspace{0.02cm}\bar{\xi}^{\,\prime(k)}\hspace{0.02cm}|\,
\hat{A}\hspace{0.03cm}a^{-}_{n}|\,\xi^{(k - 1)}\hspace{0.02cm}\rangle
=
\langle\hspace{0.02cm}\bar{\xi}^{\,\prime(k)}\hspace{0.02cm}|\,
\hat{A}\hspace{0.02cm}|\,\xi^{(k - 1)}\hspace{0.02cm}\rangle\hspace{0.03cm}\xi^{(k - 1)}_{n}.
\label{eq:7q}
\end{equation}
A similar term with the creation operator $a_{n}^{+}$ has somewhat a more complicated structure, since
\[
\hat{A}\hspace{0.03cm}a^{+}_{n} = a^{+}_{n}\hspace{0.01cm}\hat{A}
+ [\hspace{0.03cm}\hat{A}, a^{+}_{n}\hspace{0.02cm}]
\]
and therefore
\begin{equation}
\langle\hspace{0.02cm}\bar{\xi}^{\,\prime(k)}\hspace{0.02cm}|\,
\hat{A}\hspace{0.03cm}a^{+}_{n}|\,\xi^{(k - 1)}\hspace{0.02cm}\rangle
=
\label{eq:7w}
\end{equation}
\[
= \bar{\xi}^{\,\prime(k)}_{n}
\langle\hspace{0.02cm}\bar{\xi}^{\,\prime(k)}\hspace{0.02cm}|\,
\hat{A}\hspace{0.02cm}|\,\xi^{(k - 1)}\hspace{0.02cm}\rangle
-
\beta\,\langle\hspace{0.02cm}\bar{\xi}^{\,\prime(k)}\hspace{0.02cm}|\,
[\hspace{0.03cm}a^{\phantom{\pm}\!}_{0}, a^{+}_{n}\hspace{0.02cm}]\hspace{0.01cm}|\,\xi^{(k - 1)}\hspace{0.02cm}\rangle
+
\gamma\,\langle\hspace{0.02cm}\bar{\xi}^{\,\prime(k)}\hspace{0.02cm}|\,
[\hspace{0.03cm}a^{2}_{0}\hspace{0.03cm}, a^{+}_{n}\hspace{0.02cm}]\hspace{0.01cm}|\,
\xi^{(k - 1)}\hspace{0.02cm}\rangle.
\]
Here, we have taken into account the representation of the operator $\hat{A}$ in the form (\ref{eq:2p}) with the replacement $\omega \rightarrow -\hspace{0.02cm}a_0$. Therefore, we are confronted by the task of deriving matrix elements of the commutators $[\hspace{0.03cm}a^{\phantom{+}\!}_{0}, a^{+}_{n}\hspace{0.02cm}]$ and
$[\hspace{0.03cm}a^{2}_{0}\hspace{0.03cm}, a^{+}_{n}\hspace{0.02cm}]$.
Let us consider the first of them.\\
\indent By virtue of the representation of the operator $a_0$, Eq.\,(\ref{eq:5o}), we have
\begin{equation}
[\hspace{0.03cm}a^{\phantom{+}\!}_{0}, a^{+}_{n}\hspace{0.02cm}]
=
-\frac{1}{4}\,\Bigl(\,[\hspace{0.03cm}\{L^{\phantom{+}\!}_{12},M^{\phantom{+}\!}_{12}\}, a^{+}_{n}\hspace{0.02cm}]
+
[\hspace{0.03cm}\{N^{\phantom{+}\!}_{12},N^{\phantom{+}\!}_{21}\}, a^{+}_{n}\hspace{0.02cm}]
-
2\hspace{0.04cm}[\hspace{0.03cm}N^{\phantom{+}\!\!}_{1}N^{\phantom{+}\!}_{2}, a^{+}_{n}\hspace{0.02cm}]\hspace{0.01cm}
\Bigr).
\label{eq:7e}
\end{equation}
By using the operator identity
\[
[\hspace{0.03cm}\{A,B\hspace{0.02cm}\},C\hspace{0.03cm}]
=
\{A,[B,C\hspace{0.03cm}]\hspace{0.02cm}\} +  \{B,[\hspace{0.02cm}A,C\hspace{0.02cm}]\hspace{0.02cm}\}
\]
and the commutation rules (\ref{eq:5p}), it is not difficult to obtain a more simple form of the commutators on the right-hand side (\ref{eq:7e})
\begin{equation}
\begin{split}
&[\hspace{0.03cm}\{L^{\phantom{+}\!}_{12},M^{\phantom{+}\!}_{12}\}, a^{+}_{n}\hspace{0.02cm}]
=
2\hspace{0.02cm}L^{\phantom{+}\!}_{12}\hspace{0.03cm}
(\hspace{0.03cm}\delta^{\phantom{-}}_{n2}\hspace{0.03cm}a^{-}_{1}
- \delta^{\phantom{-}}_{n1}\hspace{0.03cm}a^{-}_{2}\hspace{0.03cm})
+
(\hspace{0.03cm}\delta^{\phantom{+}}_{n2}\hspace{0.03cm}a^{+}_{2}
+ \delta^{\phantom{+}}_{n1}\hspace{0.03cm}a^{+}_{1}\hspace{0.03cm}), \\[1ex]
&[\hspace{0.03cm}\{N^{\phantom{+}\!}_{12}, N^{\phantom{+}\!}_{21}\}, a^{+}_{n}\hspace{0.02cm}]
=
2\hspace{0.03cm}\delta^{\phantom{+}}_{n2}\hspace{0.03cm}a^{+}_{1}N^{\phantom{+}\!}_{21}
+
2\hspace{0.03cm}\delta^{\phantom{+}}_{n1}\hspace{0.03cm}a^{+}_{2}N^{\phantom{+}\!}_{12}
+
(\hspace{0.03cm}\delta^{\phantom{+}}_{n2}\hspace{0.03cm}a^{+}_{2}
+ \delta^{\phantom{+}}_{n1}\hspace{0.03cm}a^{+}_{1}\hspace{0.03cm}), \\[1ex]
&[\hspace{0.03cm}N^{\phantom{+}\!\!}_{1}N^{\phantom{+}\!}_{2}, a^{+}_{n}\hspace{0.02cm}]
=
\delta^{\phantom{+}}_{n2}\hspace{0.03cm}a^{+}_{2}N^{\phantom{+}\!}_{1} +
\delta^{\phantom{+}}_{n1}\hspace{0.03cm}a^{+}_{1}N^{\phantom{+}\!}_{2}.
\end{split}
\label{eq:7t}
\end{equation}
\indent We need the matrix elements of the generators $L_{12},\,M_{12}, \ldots ,$ which can be easily obtained from their definitions (\ref{eq:5i}):
\begin{equation}
\begin{array}{llll}
&\langle\hspace{0.02cm}\bar{\xi}^{\,\prime}\hspace{0.02cm}|\,
L_{12}|\,\xi\hspace{0.02cm}\rangle
=
\biggl(\displaystyle\frac{1}{2}\;[\hspace{0.03cm}\bar{\xi}^{\,\prime}_{1}, \bar{\xi}^{\,\prime}_{2}\hspace{0.03cm}]\biggr)
\langle\hspace{0.02cm}\bar{\xi}^{\,\prime}\hspace{0.02cm}|\,\xi\hspace{0.02cm}\rangle,
\quad\quad
&\langle\hspace{0.02cm}\bar{\xi}^{\,\prime}\hspace{0.02cm}|\,
M_{12}|\,\xi\hspace{0.02cm}\rangle
=
\biggl(\displaystyle\frac{1}{2}\;[\hspace{0.03cm}\xi^{\phantom{\prime}}_{1}, \xi^{\phantom{\prime}}_{2}\hspace{0.03cm}]\biggr)
\langle\hspace{0.02cm}\bar{\xi}^{\,\prime}\hspace{0.02cm}|\,\xi\hspace{0.02cm}\rangle, \\[1.5ex]
&\langle\hspace{0.02cm}\bar{\xi}^{\,\prime}\hspace{0.02cm}|\,
N_{12}|\,\xi\hspace{0.02cm}\rangle
=
\biggl(\displaystyle\frac{1}{2}\;[\hspace{0.03cm}\bar{\xi}^{\,\prime}_{1}, \xi^{\phantom{\prime}}_{2}\hspace{0.03cm}]\biggr)
\langle\hspace{0.02cm}\bar{\xi}^{\,\prime}\hspace{0.02cm}|\,\xi\hspace{0.02cm}\rangle,
\quad\quad
&\langle\hspace{0.02cm}\bar{\xi}^{\,\prime}\hspace{0.02cm}|\,
N_{21}|\,\xi\hspace{0.02cm}\rangle
=
\biggl(\displaystyle\frac{1}{2}\;[\hspace{0.03cm}\bar{\xi}^{\,\prime}_{2}, \xi^{\phantom{\prime}}_{1}\hspace{0.03cm}]\biggr)
\langle\hspace{0.02cm}\bar{\xi}^{\,\prime}\hspace{0.02cm}|\,\xi\hspace{0.02cm}\rangle, \\[1.5ex]
&\langle\hspace{0.02cm}\bar{\xi}^{\,\prime}\hspace{0.02cm}|\,
N_{1}|\,\xi\hspace{0.02cm}\rangle
=
\biggl\{\biggl(\displaystyle\frac{1}{2}\;[\hspace{0.03cm}\bar{\xi}^{\,\prime}_{1}, \xi^{\phantom{\prime}}_{1}\hspace{0.03cm}]\biggr) - 1\biggr\}\hspace{0.03cm}
\langle\hspace{0.02cm}\bar{\xi}^{\,\prime}\hspace{0.02cm}|\,\xi\hspace{0.02cm}\rangle,
\quad\quad
&\langle\hspace{0.02cm}\bar{\xi}^{\,\prime}\hspace{0.02cm}|\,
N_{2}|\,\xi\hspace{0.02cm}\rangle
=
\biggl\{\biggl(\displaystyle\frac{1}{2}\;[\hspace{0.03cm}\bar{\xi}^{\,\prime}_{2}, \xi^{\phantom{\prime}}_{2}\hspace{0.03cm}]\biggr) - 1\biggr\}\hspace{0.03cm}
\langle\hspace{0.02cm}\bar{\xi}^{\,\prime}\hspace{0.02cm}|\,\xi\hspace{0.02cm}\rangle.
\end{array}
\label{eq:7y}
\end{equation}
Hereinafter, for the sake of simplification of the notations, we omit the iteration numbers $(k)$ and $(k-1)$ of $\bar{\xi}^{\prime}$ and $\xi$. Substituting (\ref{eq:7t}) into (\ref{eq:7e}) and taking into account (\ref{eq:7y}), we obtain the desired matrix element
\begin{equation}
\langle\hspace{0.02cm}\bar{\xi}^{\,\prime}\hspace{0.02cm}|\,
[\hspace{0.03cm}a^{\phantom{+}\!}_{0}, a^{+}_{n}\hspace{0.02cm}]|\,\xi\hspace{0.02cm}\rangle
=
-\frac{1}{2}\,\biggl\{\!\biggl(\displaystyle\frac{1}{2}\;[\hspace{0.03cm}\bar{\xi}^{\,\prime}_{1}, \bar{\xi}^{\,\prime}_{2}\hspace{0.03cm}]\biggr)
(\hspace{0.03cm}\delta^{\phantom{-}}_{n2}\hspace{0.03cm}\xi_{1}
- \delta^{\phantom{-}}_{n1}\hspace{0.03cm}\xi_{2}\hspace{0.03cm})
+
\delta^{\phantom{-}}_{n1}\hspace{0.03cm}\bar{\xi}^{\,\prime}_{2}\hspace{0.01cm}
\biggl(\displaystyle\frac{1}{2}\;[\hspace{0.03cm}\bar{\xi}^{\,\prime}_{1}, \xi^{\phantom{\prime}}_{2}\hspace{0.03cm}]\biggr)\hspace{0.03cm}
+
\label{eq:7u}
\end{equation}
\[
+\, \delta^{\phantom{-}}_{n2}\hspace{0.03cm}\bar{\xi}^{\,\prime}_{1}\hspace{0.01cm}
\biggl(\displaystyle\frac{1}{2}\;[\hspace{0.03cm}\bar{\xi}^{\,\prime}_{2}, \xi^{\phantom{\prime}}_{1}\hspace{0.03cm}]\biggr)
-
\delta^{\phantom{-}}_{n2}\hspace{0.03cm}\bar{\xi}^{\,\prime}_{2}\hspace{0.01cm}
\biggl(\displaystyle\frac{1}{2}\;[\hspace{0.03cm}\bar{\xi}^{\,\prime}_{1}, \xi^{\phantom{\prime}}_{1}\hspace{0.03cm}]\biggr)
-
\delta^{\phantom{-}}_{n1}\hspace{0.03cm}\bar{\xi}^{\,\prime}_{1}\hspace{0.01cm}
\biggl(\displaystyle\frac{1}{2}\;[\hspace{0.03cm}\bar{\xi}^{\,\prime}_{2}, \xi^{\phantom{\prime}}_{2}\hspace{0.03cm}]\biggr)
+
2\hspace{0.03cm}(\hspace{0.03cm}\delta^{\phantom{+}}_{n2}\hspace{0.03cm}
\bar{\xi}^{\,\prime}_{2}
+ \delta^{\phantom{+}}_{n1}\hspace{0.03cm}\bar{\xi}^{\,\prime}_{1}\hspace{0.03cm})
\hspace{0.03cm}\!\biggr\}\hspace{0.03cm}
\langle\hspace{0.02cm}\bar{\xi}^{\,\prime}\hspace{0.02cm}|\,\xi\hspace{0.02cm}\rangle.
\]
\indent The expression can be presented in a more compact and visual form. For this purpose we rewrite matrix element of operator $a_{0}$ in the following form:
\begin{equation}
\langle\hspace{0.02cm}\bar{\xi}^{\,\prime}\hspace{0.02cm}|\,
a_{0}|\,\xi\hspace{0.02cm}\rangle
=
\Omega\,\langle\hspace{0.02cm}\bar{\xi}^{\,\prime}\hspace{0.02cm}|\,\xi\hspace{0.02cm}\rangle,
\label{eq:7i}
\end{equation}
where in accordance with (\ref{eq:5f}) we have
\begin{equation}
\begin{split}
\Omega \equiv \Omega\hspace{0.03cm}(\bar{\xi}^{\,\prime}, \xi)
=
-\frac{1}{2}\,\biggl\{\!
&\biggl(\displaystyle\frac{1}{2}\;[\hspace{0.03cm}\bar{\xi}^{\,\prime}_{1}, \bar{\xi}^{\,\prime}_{2}\hspace{0.03cm}]\biggr)\!
\biggl(\,\displaystyle\frac{1}{2}\;[\hspace{0.03cm}\xi^{\phantom{\prime}}_{1}, \xi^{\phantom{\prime}}_{2}\hspace{0.03cm}]\biggr)
+
\biggl(\displaystyle\frac{1}{2}\;[\hspace{0.03cm}\bar{\xi}^{\,\prime}_{1}, \xi^{\phantom{\prime}}_{2}\hspace{0.03cm}]\biggr)\!
\biggl(\displaystyle\frac{1}{2}\;[\hspace{0.03cm}\bar{\xi}^{\,\prime}_{2}, \xi^{\phantom{\prime}}_{1}\hspace{0.03cm}]\biggr) - \\[1ex]
-\, &\biggl(\displaystyle\frac{1}{2}\;[\hspace{0.03cm}\bar{\xi}^{\,\prime}_{1}, \xi^{\phantom{\prime}}_{1}\hspace{0.03cm}]\biggr)\!
\biggl(\displaystyle\frac{1}{2}\;[\hspace{0.03cm}\bar{\xi}^{\,\prime}_{2}, \xi^{\phantom{\prime}}_{2}\hspace{0.03cm}]\biggr)
+
2\biggl(\displaystyle\frac{1}{2}\;[\hspace{0.03cm}\bar{\xi}^{\,\prime}_{1}, \xi^{\phantom{\prime}}_{1}\hspace{0.03cm}]
+
\displaystyle\frac{1}{2}\;[\hspace{0.03cm}\bar{\xi}^{\,\prime}_{2}, \xi^{\phantom{\prime}}_{2}\hspace{0.03cm}] - 1\biggr)\!
\biggr\}.
\end{split}
\label{eq:7o}
\end{equation}
Let us take the derivative of the function $\Omega$ with respect to $\xi_{n}$ by making use of the rules of differentiation \eqref{ap:B6} and \eqref{ap:B7}
\begin{equation}
\frac{\partial\hspace{0.04cm}\Omega}{\partial\hspace{0.02cm}\xi_{n}}
=
-\frac{1}{2}\,\biggl\{\!\biggl(\displaystyle\frac{1}{2}\;[\hspace{0.03cm}\bar{\xi}^{\,\prime}_{1}, \bar{\xi}^{\,\prime}_{2}\hspace{0.03cm}]\biggr)
(\hspace{0.03cm}\delta^{\phantom{-}}_{n1}\hspace{0.03cm}\xi_{2}
- \delta^{\phantom{-}}_{n2}\hspace{0.03cm}\xi_{1}\hspace{0.03cm})\,-
\label{eq:7p}
\end{equation}
\[
-\: \frac{1}{2}\:\bar{\xi}^{\,\prime}_{2}\,
[\hspace{0.03cm}\bar{\xi}^{\,\prime}_{1}, (\delta^{\phantom{-}}_{n1}\hspace{0.03cm}\xi^{\phantom{\prime}}_{2} -
\delta^{\phantom{-}}_{n2}\hspace{0.03cm}\xi^{\phantom{\prime}}_{1})\hspace{0.03cm}]
+
\frac{1}{2}\:\bar{\xi}^{\,\prime}_{1}\,
[\hspace{0.03cm}\bar{\xi}^{\,\prime}_{2}, (\delta^{\phantom{-}}_{n1}\hspace{0.03cm}\xi^{\phantom{\prime}}_{2} -
\delta^{\phantom{-}}_{n2}\hspace{0.03cm}\xi^{\phantom{\prime}}_{1})\hspace{0.03cm}]
-
2\hspace{0.03cm}(\hspace{0.03cm}\delta^{\phantom{+}}_{n1}\hspace{0.03cm}
\bar{\xi}^{\,\prime}_{1}
+ \delta^{\phantom{+}}_{n2}\hspace{0.03cm}\bar{\xi}^{\,\prime}_{2}\hspace{0.03cm})
\hspace{0.04cm}\!\biggr\}.
\]
Comparing the last expression with (\ref{eq:7u}), we obtain that
\begin{equation}
\langle\hspace{0.02cm}\bar{\xi}^{\,\prime}\hspace{0.02cm}|\,
[\hspace{0.03cm}a^{\phantom{+}}_{0}\!, a^{+}_{n}\hspace{0.02cm}]|\,\xi\hspace{0.02cm}\rangle
=
-\biggl(\frac{\partial\hspace{0.04cm}\Omega}{\partial\hspace{0.02cm}\xi_{n}}\biggr)
\hspace{0.02cm}
\langle\hspace{0.02cm}\bar{\xi}^{\,\prime}\hspace{0.02cm}|\,\xi\hspace{0.02cm}\rangle.
\label{eq:7a}
\end{equation}
Similar reasoning for the commutator $[\hspace{0.03cm}a^{\phantom{-}}_{0}\!, a^{-}_{n}\hspace{0.02cm}]$ leads us to the representation of the corresponding matrix element
\begin{equation}
\langle\hspace{0.02cm}\bar{\xi}^{\,\prime}\hspace{0.02cm}|\,
[\hspace{0.03cm}a^{\phantom{-}\!}_{0}, a^{-}_{n}\hspace{0.02cm}]|\,\xi\hspace{0.02cm}\rangle
=
-\biggl(\frac{\partial\hspace{0.04cm}\Omega}
{\partial\hspace{0.02cm}\bar{\xi}^{\,\prime}_{n}}\biggr) \hspace{0.02cm}
\langle\hspace{0.02cm}\bar{\xi}^{\,\prime}\hspace{0.02cm}|\,\xi\hspace{0.02cm}\rangle,
\label{eq:7s}
\end{equation}
where
\begin{equation}
\frac{\partial\hspace{0.04cm}\Omega}{\partial\hspace{0.02cm}\bar{\xi}^{\,\prime}_{n}}
=
-\frac{1}{2}\,\biggl\{
(\hspace{0.03cm}\delta^{\phantom{-}}_{n1}\hspace{0.04cm}\bar{\xi}^{\,\prime}_{2}
- \delta^{\phantom{-}}_{n2}\hspace{0.04cm}\bar{\xi}^{\,\prime}_{1}\hspace{0.03cm})
\biggl(\displaystyle\frac{1}{2}\;
[\hspace{0.03cm}\xi^{\phantom{\prime}}_{1}, \xi^{\phantom{\prime}}_{2}\hspace{0.03cm}]\biggr)
-
\label{eq:7d}
\end{equation}
\[
-\,
\frac{1}{2}\;[\hspace{0.03cm}
(\hspace{0.03cm}\delta^{\phantom{-}}_{n1}\hspace{0.04cm}\bar{\xi}^{\,\prime}_{2}
- \delta^{\phantom{-}}_{n2}\hspace{0.04cm}\bar{\xi}^{\,\prime}_{1}\hspace{0.03cm}), \xi^{\phantom{\prime}}_{2}\hspace{0.03cm}]\,\xi^{\phantom{\prime}}_{1}
\,+
\frac{1}{2}\;[\hspace{0.03cm}
(\hspace{0.03cm}\delta^{\phantom{-}}_{n1}\hspace{0.03cm}\bar{\xi}^{\,\prime}_{2}
- \delta^{\phantom{-}}_{n2}\hspace{0.04cm}\bar{\xi}^{\,\prime}_{1}\hspace{0.03cm}), \xi^{\phantom{\prime}}_{1}\hspace{0.03cm}]\,\xi^{\phantom{\prime}}_{2}
+
2\hspace{0.03cm}(\hspace{0.03cm}\delta^{\phantom{+}}_{n1}\hspace{0.04cm}
\xi^{\phantom{\prime}}_{1}
+ \delta^{\phantom{+}}_{n2}\hspace{0.03cm}\xi^{\phantom{\prime}}_{2}\hspace{0.03cm})
\hspace{0.03cm}\!\biggr\}.
\]
\indent Further, in the paper \cite{markov_2020} it is shown that the matrix element  $\langle\hspace{0.02cm}\bar{\xi}^{\,\prime}\hspace{0.02cm}|\,[\hspace{0.03cm}a^{2}_{0},  a^{+}_{n}\hspace{0.02cm}]|\,\xi\hspace{0.02cm}\rangle$ has the following form:
\begin{equation}
\langle\hspace{0.02cm}\bar{\xi}^{\,\prime}\hspace{0.02cm}|\,
[\hspace{0.03cm}a^{2}_{0}\hspace{0.02cm}, a^{+}_{n}\hspace{0.02cm}]|\,\xi\hspace{0.02cm}\rangle
=
\label{eq:7f}
\end{equation}
\[
\begin{split}
&=\bar{\xi}^{\,\prime}_{n}\hspace{0.03cm}\biggl[\,\delta_{n1}\hspace{0.02cm}
\biggl\{- 1 - 2\hspace{0.03cm}\biggl(\displaystyle\frac{1}{2}\;[\hspace{0.01cm}\bar{\xi}^{\,\prime}_{1}, \xi^{\phantom{\prime}}_{1}\hspace{0.03cm}] - 1\biggr)
+ 2\hspace{0.03cm}\biggl[\hspace{0.03cm}\biggl(\displaystyle\frac{1}{2}\;[\hspace{0.03cm}
\bar{\xi}^{\,\prime}_{2}, \xi^{\phantom{\prime}}_{2}\hspace{0.03cm}]\biggr)^{\!\!2}
-
\biggl(\displaystyle\frac{1}{2}\;[\hspace{0.03cm}\bar{\xi}^{\,\prime}_{2}, \xi^{\phantom{\prime}}_{2}\hspace{0.03cm}]\biggr)
+ 1\,\biggr]  \hspace{0.03cm}+ \\[1ex]
&+\hspace{0.03cm} 4\hspace{0.03cm}\biggl(\displaystyle\frac{1}{2}\;[\hspace{0.03cm}\bar{\xi}^{\,\prime}_{1}, \xi^{\phantom{\prime}}_{1}\hspace{0.03cm}] - 1\biggr)
\biggl[\,\biggl(\displaystyle\frac{1}{2}\;[\hspace{0.03cm}\bar{\xi}^{\,\prime}_{2}, \xi^{\phantom{\prime}}_{2}\hspace{0.03cm}]\biggr)^{\!\!2}
-
\biggl(\displaystyle\frac{1}{2}\;[\hspace{0.03cm}\bar{\xi}^{\,\prime}_{2}, \xi^{\phantom{\prime}}_{2}\hspace{0.03cm}]\biggr)
+ 1\,\biggr]\biggr\}
+ (1\rightleftarrows 2)\,\biggr]\hspace{0.02cm}
\langle\hspace{0.02cm}\bar{\xi}^{\,\prime}\hspace{0.02cm}|\,\xi\hspace{0.02cm}\rangle.
\end{split}
\]
\indent Let us present the expression (\ref{eq:7f}) in the form similar to the form (\ref{eq:7a}) for the matrix element of commutator the $[\hspace{0.03cm}a^{\phantom{+}}_{0}\!, a^{+}_{n}\hspace{0.02cm}]$. For this purpose, we write out the matrix element of the operator $a_{0}^{2}$ in the form
\begin{equation}
\langle\hspace{0.02cm}\bar{\xi}^{\,\prime}\hspace{0.02cm}|\,
a^{2}_{0}\,|\,\xi\hspace{0.02cm}\rangle = \widetilde{\Omega} \,
\langle\hspace{0.02cm}\bar{\xi}^{\,\prime}\hspace{0.02cm}|\,\xi\hspace{0.02cm}\rangle,
\label{eq:7g}
\end{equation}
where in accordance with (\ref{eq:6y}), we have
\begin{equation}
\widetilde{\Omega} = \widetilde{\Omega}\hspace{0.03cm}(\bar{\xi}^{\,\prime},\xi) =
\label{eq:7h}
\end{equation}
\[
= 1 - \sum\limits^{2}_{k\hspace{0.015cm}=\hspace{0.015cm}1}
\biggl[\biggl(\displaystyle\frac{1}{2}\;[\hspace{0.03cm}\bar{\xi}^{\,\prime}_{k}, \xi^{\phantom{\prime}}_{k}\hspace{0.02cm}]\biggr)^{\!\!2\,} -
\displaystyle\frac{1}{2}\;[\hspace{0.03cm}\bar{\xi}^{\,\prime}_{k}, \xi^{\phantom{\prime}}_{k}\hspace{0.02cm}]
+ 1\biggr]\,
+ 2\hspace{0.02cm}\prod\limits^{2}_{k\hspace{0.015cm}=\hspace{0.015cm}1}
\biggl[\biggl(\displaystyle\frac{1}{2}\;[\hspace{0.03cm}\bar{\xi}^{\,\prime}_{k}, \xi^{\phantom{\prime}}_{k}\hspace{0.02cm}]\biggr)^{\!\!2\,} -
\displaystyle\frac{1}{2}\;[\hspace{0.03cm}\bar{\xi}^{\,\prime}_{k}, \xi^{\phantom{\prime}}_{k}\hspace{0.02cm}] + 1\biggr].
\]
By a direct calculation, using the formulas of differentiation \eqref{ap:B6} and \eqref{ap:B7}, it is easy to verify that  the following relation
\begin{equation}
\langle\hspace{0.02cm}\bar{\xi}^{\,\prime}\hspace{0.02cm}|\,
[\hspace{0.03cm}a^{2}_{0}\hspace{0.03cm}, a^{+}_{n}\hspace{0.02cm}]|\,\xi\hspace{0.02cm}\rangle
=
-\biggl(\frac{\partial\hspace{0.04cm}\widetilde{\Omega}}{\partial\hspace{0.02cm}\xi_{n}}\biggr)
\hspace{0.02cm}
\langle\hspace{0.02cm}\bar{\xi}^{\,\prime}\hspace{0.02cm}|\,\xi\hspace{0.02cm}\rangle
\label{eq:7j}
\end{equation}
is true. The same reasoning leads to
\begin{equation}
\langle\hspace{0.02cm}\bar{\xi}^{\,\prime}\hspace{0.02cm}|\,
[\hspace{0.03cm}a^{2}_{0}\hspace{0.03cm}, a^{-}_{n}\hspace{0.02cm}]|\,\xi\hspace{0.02cm}\rangle
=
-\biggl(\frac{\partial\hspace{0.04cm}\widetilde{\Omega}}
{\partial\hspace{0.02cm}\bar{\xi}^{\,\prime}_{n}}\biggr) \hspace{0.02cm}
\langle\hspace{0.02cm}\bar{\xi}^{\,\prime}\hspace{0.02cm}|\,\xi\hspace{0.02cm}\rangle.
\label{eq:7k}
\end{equation}
\indent Let us return to the matrix element (\ref{eq:7w}). We write the matrix element of the operator $\hat{A}$ in a form similar to the form of expressions (\ref{eq:7i}) and (\ref{eq:7g}):
\begin{equation}
\langle\hspace{0.02cm}\bar{\xi}^{\,\prime}\hspace{0.02cm}|\,
\hat{A}\,|\,\xi\hspace{0.02cm}\rangle = {\cal A}\,
\langle\hspace{0.02cm}\bar{\xi}^{\,\prime}\hspace{0.02cm}|\,\xi\hspace{0.02cm}\rangle.
\label{eq:7l}
\end{equation}
The function ${\cal A}={\cal A}(\bar{\xi}^{\,\prime},\xi)$ can be written out based on the expression (\ref{eq:4e}) (with the replacement $\hat{\omega} \rightarrow -\hspace{0.02cm} a_{0}$) and with allowance made for (\ref{eq:7i}), (\ref{eq:7o}) and (\ref{eq:7g}), (\ref{eq:7h}). This will be done in Part II, where we will consider in detail a question of a connection between the operator $a_{0}^{2}$ defined by the expression (\ref{ap:A7}) and the {\it square} of the operator $a_{0}$ (\hspace{0.02cm}i.e. $(a_{0})^{2}\equiv a_{0}\cdot a_{0}$) defined by the expression (\ref{eq:5o}).\\
\indent For the remaining two terms in (\ref{eq:7w}) we use the representations (\ref{eq:7a}) and (\ref{eq:7j}), correspondingly. As a result, instead of (\ref{eq:7w}), we have
\begin{equation}
\langle\hspace{0.02cm}\bar{\xi}^{\,\prime}\hspace{0.02cm}|\,
\hat{A}\hspace{0.03cm}a^{+}_{n}|\,\xi\hspace{0.02cm}\rangle
=
\Bigl(\hspace{0.02cm}\bar{\xi}^{\,\prime}_{n}\hspace{0.03cm}{\cal A}
- \frac{\partial{\cal A}}{\partial\xi_{n}}\hspace{0.02cm}\Bigr)\hspace{0.02cm}
\langle\hspace{0.02cm}\bar{\xi}^{\,\prime}\hspace{0.02cm}|\,\xi\hspace{0.02cm}\rangle.
\label{eq:7z}
\end{equation}
Follow the same procedure, we can write out the matrix element for the product $a^{-}_{n}\hspace{0.02cm}\hat{A}$
\begin{equation}
\langle\hspace{0.02cm}\bar{\xi}^{\,\prime}\hspace{0.02cm}|\,
a^{-}_{n}\hspace{0.03cm}\hat{A}|\,\xi\hspace{0.02cm}\rangle
=
\Bigl(\hspace{0.02cm}\xi_{n}\hspace{0.03cm}{\cal A}
- \frac{\partial{\cal A}}{\partial\bar{\xi}^{\,\prime}_{n}}\hspace{0.02cm}\Bigr)\hspace{0.02cm}
\langle\hspace{0.02cm}\bar{\xi}^{\,\prime}\hspace{0.02cm}|\,\xi\hspace{0.02cm}\rangle.
\label{eq:7x}
\end{equation}
The last two expressions will be used in the following section. In the accepted notations the matrix element (\ref{eq:7q}) is rewritten in the form
\[
\langle\hspace{0.02cm}\bar{\xi}^{\,\prime}\hspace{0.02cm}|\,
\hat{A}\hspace{0.03cm}a^{-}_{n}|\,\xi\hspace{0.02cm}\rangle = \xi_{n}\hspace{0.03cm}{\cal A}\,
\langle\hspace{0.02cm}\bar{\xi}^{\,\prime}\hspace{0.02cm}|\,\xi\hspace{0.02cm}\rangle.
\]
The absence of the term with derivative in the last expression in comparison with (\ref{eq:7x}) is connected with the fact that operators $\hat{A}$ and $ a_{n}^{-}$ are not commutative.


\section{Matrix elements of the product 
\texorpdfstring{$\hat{A}\hspace{0.03cm}[\hspace{0.03cm}a^{ }_{0}, a^{\pm}_{n}\hspace{0.02cm}]$}{a0a2}}
\setcounter{equation}{0}
\label{section_8}

Now we proceed to analysis of the remaining matrix elements in the initial expression (\ref{eq:4w}), namely, to analysis of  $\langle\hspace{0.02cm}\bar{\xi}^{\,\prime(k)}\hspace{0.02cm}|\hspace{0.02cm}
\hat{A}\hspace{0.03cm}[\hspace{0.03cm}a^{\phantom{\pm}\!}_{0},a^{\pm}_{n}\hspace{0.02cm}]
\hspace{0.02cm}|\,\xi^{(k - 1)}\hspace{0.02cm}\rangle$. As in the previous section, for brevity the indices $(k)$ and $(k-1)$, namely the iteration numbers are omitted.\\
\indent We need the following representations for the commutators $[\hspace{0.03cm}a^{\phantom{-}\!}_{0},a^{-}_{n}\hspace{0.02cm}]$ and $[\hspace{0.03cm}a^{\phantom{+}\!}_{0},a^{+}_{n}\hspace{0.02cm}]$:
\begin{align}
[\hspace{0.03cm}a^{\phantom{-}\!}_{0}, a^{-}_{n}\hspace{0.02cm}]
=
-\frac{1}{2}\,\Bigl\{&\hspace{0.03cm}
(\hspace{0.03cm}\delta^{\phantom{+}}_{n2}\hspace{0.03cm}a^{+}_{1} - \delta^{\phantom{+}}_{n1}\hspace{0.03cm}a^{+}_{2}\hspace{0.03cm})M^{\phantom{+}\!}_{12}
-
(\hspace{0.03cm}\delta_{n1}\hspace{0.02cm}N^{\phantom{+}\!}_{21}\hspace{0.03cm}a^{-}_{2} +
\hspace{0.03cm}\delta_{n2}\hspace{0.02cm}N^{\phantom{+}\!}_{12}\hspace{0.03cm}a^{-}_{1}\hspace{0.03cm})\, +
\label{eq:8q} \\[1ex]
&+
(\hspace{0.03cm}\delta_{n2}\hspace{0.03cm}N^{\phantom{+}\!\!}_{1}\hspace{0.03cm}a^{-}_{2} +
\delta_{n1}\hspace{0.03cm}N^{\phantom{+}\!\!}_{2}\hspace{0.03cm}a^{-}_{1}\hspace{0.03cm})
-
(\hspace{0.03cm}\delta^{\phantom{-}}_{n2}\hspace{0.03cm}a^{-}_{2}
+ \delta^{\phantom{-}}_{n1}\hspace{0.03cm}a^{-}_{1}\hspace{0.03cm})
\!\hspace{0.03cm}\Bigr\}, \notag
\end{align}
\begin{align}
[\hspace{0.03cm}a^{\phantom{+}\!}_{0}, a^{+}_{n}\hspace{0.02cm}]
=
-\frac{1}{2}\,\Bigl\{&L^{\phantom{+}\!}_{12}\hspace{0.03cm}
(\hspace{0.03cm}\delta^{\phantom{-}}_{n2}\hspace{0.03cm}a^{-}_{1} - \delta^{\phantom{-}}_{n1}\hspace{0.03cm}a^{-}_{2}\hspace{0.03cm})
+
(\hspace{0.03cm}\delta_{n2}\hspace{0.03cm}a^{+}_{1}N^{\phantom{+}\!}_{21} +
\hspace{0.03cm}\delta_{n1}\hspace{0.03cm}a^{+}_{2}N^{\phantom{+}\!\!}_{12}\hspace{0.03cm})\, -
\label{eq:8w} \\[1ex]
&-
(\hspace{0.03cm}\delta_{n2}\hspace{0.03cm}a^{+}_{2}N^{\phantom{+}\!\!}_{1} +
\delta_{n1}\hspace{0.03cm}a^{+}_{1}N^{\phantom{+}\!\!}_{2}\hspace{0.03cm})
+
(\hspace{0.03cm}\delta^{\phantom{+}}_{n2}\hspace{0.03cm}a^{+}_{2}
+ \delta^{\phantom{+}}_{n1}\hspace{0.03cm}a^{+}_{1}\hspace{0.03cm})
\!\hspace{0.03cm}\Bigr\}. \notag
\end{align}
A proof of the second representation was given in section \ref{section_7}, Eqs.\,(\ref{eq:7e}) and (\ref{eq:7t}), the first one is proved in a similar way. Further we consider action of the commutator (\ref{eq:8q}) on the parafermion coherent state
\begin{align}
[\hspace{0.03cm}a^{\phantom{-}\!}_{0}, a^{-}_{n}\hspace{0.02cm}]|\,\xi\hspace{0.02cm}\rangle
=
-\frac{1}{2}\,\biggl\{\hspace{0.03cm}&(\hspace{0.03cm}\delta^{\phantom{+}}_{n2}
\hspace{0.03cm}a^{+}_{1} - \delta^{\phantom{+}}_{n1}\hspace{0.03cm}a^{+}_{2}\hspace{0.03cm})
\biggl(\displaystyle\frac{1}{2}\;[\hspace{0.03cm}\xi^{\phantom{\prime}}_{1}, \xi^{\phantom{\prime}}_{2}\hspace{0.03cm}]\biggr)
-
(\hspace{0.03cm}\delta^{\phantom{+}\!}_{n1}\hspace{0.02cm}N^{\phantom{+}\!}_{21}\hspace{0.03cm}\xi^{\phantom{\prime}}_{2} +
\hspace{0.03cm}\delta_{n2}\hspace{0.02cm}N^{\phantom{+}\!}_{12}\hspace{0.03cm}\xi^{\phantom{\prime}}_{1}
\hspace{0.03cm})\, +
\label{eq:8e} \\[1ex]
&+
(\hspace{0.03cm}\delta^{\phantom{+}\!}_{n2}\hspace{0.03cm}N^{\phantom{+}\!\!}_{1}\hspace{0.03cm}\xi^{\phantom{\prime}}_{2} +
\delta_{n1}\hspace{0.03cm}N^{\phantom{+}\!\!}_{2}\hspace{0.03cm}\xi^{\phantom{\prime}}_{1}\hspace{0.03cm})
-
(\hspace{0.03cm}\delta^{\phantom{-}}_{n2}\hspace{0.03cm}\xi^{\phantom{\prime}}_{2}
+ \delta^{\phantom{-}}_{n1}\hspace{0.03cm}\xi^{\phantom{\prime}}_{1}\hspace{0.03cm})
\!\hspace{0.03cm}\biggr\}|\,\xi\hspace{0.02cm}\rangle. \notag
\end{align}
Action of the generators $N_{1}, N_{2}, N_{12}$ and $N_{21}$ on the coherent state has the following form:
\begin{equation}
\begin{array}{llll}
&N^{\phantom{+}\!}_{12}|\,\xi\hspace{0.02cm}\rangle
=
\biggl(\displaystyle\frac{1}{2}\;[\hspace{0.03cm}a^{+}_{1}, \xi^{\phantom{+}}_{2}]\biggr) |\,\xi\hspace{0.02cm}\rangle,
\quad\quad
&N^{\phantom{+}\!\!}_{1}|\,\xi\hspace{0.02cm}\rangle
=
\biggl(\displaystyle\frac{1}{2}\;[\hspace{0.03cm}a^{+}_{1}, \xi^{\phantom{+}}_{1}\!\hspace{0.01cm}] - 1\biggr)
|\,\xi\hspace{0.02cm}\rangle, \\[2ex]
&N^{\phantom{+}\!}_{21}|\,\xi\hspace{0.02cm}\rangle
=
\biggl(\displaystyle\frac{1}{2}\;[\hspace{0.03cm}a^{+}_{2}, \xi^{\phantom{+}}_{1}]\biggr) |\,\xi\hspace{0.02cm}\rangle,
\quad\quad
&N^{\phantom{+}\!\!}_{2}|\,\xi\hspace{0.02cm}\rangle
=
\biggl(\displaystyle\frac{1}{2}\;[\hspace{0.03cm}a^{+}_{2}, \xi^{\phantom{+}}_{2}\!\hspace{0.01cm}] - 1\biggr)
|\,\xi\hspace{0.02cm}\rangle.
\end{array}
\label{eq:8r}
\end{equation}
We note that the following relation is true:
\begin{equation}
[\hspace{0.03cm}a_{0}\hspace{0.03cm},\xi_{k}\hspace{0.02cm}] = 0,
\label{eq:8t}
\end{equation}
since the operator $a_0$ consists of only the commutators of the operators $a_{k}^{+}$ and $a_{k}^{-}$, and by virtue of \eqref{ap:B3} and \eqref{ap:B4}, the following relationships hold:
\[
[\hspace{0.03cm}[\hspace{0.02cm}a^{\pm}_{i},a^{\pm}_{j}\hspace{0.02cm}],
\xi^{\phantom{+}\!\!}_{k}\hspace{0.02cm}] = 0,
\quad
[\hspace{0.03cm}[\hspace{0.02cm}a^{\pm}_{i},a^{\mp}_{j}\hspace{0.02cm}],
\xi^{\phantom{+}\!\!}_{k}\hspace{0.02cm}] = 0.
\]
A trivial consequence of (\ref{eq:8t}) is the relation
\begin{equation}
[\hspace{0.03cm}\hat{A}\hspace{0.03cm},\xi_{k}\hspace{0.02cm}] = 0,
\label{eq:8y}
\end{equation}
which holds by the definition
\begin{equation}
\hat{A} = \alpha\hspace{0.02cm}\hat{I}  - \beta\hspace{0.02cm}a_{0}
+ \gamma\hspace{0.04cm}(a_{0})^{2}.
\label{eq:8u}
\end{equation}
We note once more that here in the last term we write exactly $(a_{0})^{2} \equiv a_{0}\cdot a_{0}$ to distinguish it from the symbol $a_{0}^{2}$, which we keep for the notation of Geyer's operator (\ref{ap:A7}).\\
\indent Taking into account the expressions (\ref{eq:8e}), (\ref{eq:8r}) and the relation (\ref{eq:8y}), we can present the matrix element of the product $\hat{A}\hspace{0.02cm}[\hspace{0.03cm} a^{\phantom{-}\!}_{0}, a^{-}_{n}\hspace{0.02cm}]$ as follows:
\begin{equation}
\langle\hspace{0.02cm}\bar{\xi}^{\,\prime}\hspace{0.02cm}|\hspace{0.03cm}
\hat{A}\hspace{0.02cm}[\hspace{0.03cm}a^{\phantom{-}\!}_{0}, a^{-}_{n}\hspace{0.02cm}]|\,\xi\hspace{0.02cm}\rangle
=
-\frac{1}{2}\,\biggl\{\langle\hspace{0.02cm}\bar{\xi}^{\,\prime}\hspace{0.02cm}|
\hspace{0.03cm}
\hat{A}\hspace{0.02cm}(\hspace{0.03cm}\delta^{\phantom{+}}_{n2}
\hspace{0.03cm}a^{+}_{1} - \delta^{\phantom{+}}_{n1}\hspace{0.03cm}a^{+}_{2}\hspace{0.03cm})
|\,\xi\hspace{0.02cm}\rangle
\biggl(\displaystyle\frac{1}{2}\;[\hspace{0.03cm}\xi^{\phantom{\prime}}_{1}, \xi^{\phantom{\prime}}_{2}\hspace{0.03cm}]\biggr)
\,-
\label{eq:8i}
\end{equation}
\[
-\,\frac{1}{2}\;
\delta^{\phantom{+}\!}_{n1}\hspace{0.03cm}[\hspace{0.03cm}\langle
\hspace{0.02cm}\bar{\xi}^{\,\prime}\hspace{0.02cm}|\,
\hat{A}\hspace{0.03cm}a^{+}_{2}|\,\xi\hspace{0.02cm}\rangle , \xi^{\phantom{\prime}}_{1}\hspace{0.03cm}]\hspace{0.04cm}\xi^{\phantom{\prime}}_{2}
-\frac{1}{2}\;
\delta^{\phantom{+}\!}_{n2}\hspace{0.03cm}[\hspace{0.03cm}\langle
\hspace{0.02cm}\bar{\xi}^{\,\prime}\hspace{0.02cm}|\,
\hat{A}\hspace{0.03cm}a^{+}_{1}|\,\xi\hspace{0.02cm}\rangle , \xi^{\phantom{\prime}}_{2}\hspace{0.03cm}]\hspace{0.04cm}\xi^{\phantom{\prime}}_{1}
\,+
\]
\[
+\,
\delta^{\phantom{+}\!}_{n2}\hspace{0.03cm}\biggl(\hspace{0.03cm}\frac{1}{2}\;
[\hspace{0.03cm}\langle\hspace{0.02cm}\bar{\xi}^{\,\prime}\hspace{0.02cm}|\,
\hat{A}\hspace{0.03cm}a^{+}_{1}|\,\xi\hspace{0.02cm}\rangle, \xi^{\phantom{\prime}}_{1}\hspace{0.03cm}]\hspace{0.04cm}\xi^{\phantom{\prime}}_{2}
-
\langle\hspace{0.02cm}\bar{\xi}^{\,\prime}\hspace{0.02cm}|\,
\hat{A}\hspace{0.03cm}|\,\xi\hspace{0.02cm}\rangle\hspace{0.04cm}\xi^{\phantom{\prime}}_{2}
\biggr)
+
\delta_{n1}\hspace{0.03cm}\biggl(\hspace{0.03cm}\frac{1}{2}\;
[\hspace{0.03cm}\langle\hspace{0.02cm}\bar{\xi}^{\,\prime}\hspace{0.02cm}|\,
\hat{A}\hspace{0.03cm}a^{+}_{2}|\,\xi\hspace{0.02cm}\rangle, \xi^{\phantom{\prime}}_{2}\hspace{0.03cm}]\hspace{0.04cm}\xi^{\phantom{\prime}}_{1}
-
\langle\hspace{0.02cm}\bar{\xi}^{\,\prime}\hspace{0.02cm}|\,
\hat{A}\hspace{0.03cm}|\,\xi\hspace{0.02cm}\rangle\hspace{0.04cm}\xi^{\phantom{\prime}}_{1}
\biggr)
-
\]
\[
-\,
(\hspace{0.03cm}\delta^{\phantom{+}}_{n2}
\hspace{0.03cm}\xi^{\phantom{+}}_{2} - \delta^{\phantom{+}}_{n1}\hspace{0.03cm}\xi^{\phantom{+}}_{1}\!)
\hspace{0.02cm}\langle\hspace{0.02cm}\bar{\xi}^{\,\prime}\hspace{0.02cm}|\,
\hat{A}\hspace{0.03cm}|\,\xi\hspace{0.02cm}\rangle\!\hspace{0.03cm}\biggr\}.
\]
Thus we have been able to reduce the calculation of the initial matrix element
$\langle\hspace{0.02cm}\bar{\xi}^{\,\prime}\hspace{0.02cm}|\hspace{0.03cm}\hat{A}
\hspace{0.02cm} [\hspace{0.03cm}a^{\phantom{-}\!}_{0}, a^{-}_{n}\hspace{0.02cm}]| \,\xi\hspace{0.02cm}\rangle$
to that of the matrix elements
$\langle\hspace{0.02cm}\bar{\xi}^{\,\prime}\hspace{0.02cm}|\,\hat{A}\hspace{0.03cm}|\,\xi \hspace{0.02cm}\rangle$
and
$\langle\hspace{0.02cm}\bar{\xi}^{\,\prime}\hspace{0.02cm}|\,
\hat{A}\hspace{0.03cm}a^{+}_{n}|\,\xi\hspace{0.02cm}\rangle$,
which in turn are given by (\ref{eq:7l}) and (\ref{eq:7z}), correspondingly. Collecting similar terms and recalling the definition of the derivative $\partial\hspace{0.04cm}\Omega/\partial\hspace{0.02cm} \bar{\xi}_{n}^{\prime}$, Eq.\,(\ref{eq:7d}), we can write the expression (\ref{eq:8i}) in a more compact form
\begin{equation}
\langle\hspace{0.02cm}\bar{\xi}^{\,\prime}\hspace{0.02cm}|\hspace{0.03cm}
\hat{A}\hspace{0.02cm}[\hspace{0.03cm}a^{\phantom{-}\!}_{0}, a^{-}_{n}\hspace{0.02cm}]|\,\xi\hspace{0.02cm}\rangle
=
\biggl\{-\frac{\partial\hspace{0.04cm}\Omega}
{\partial\hspace{0.02cm}\bar{\xi}^{\,\prime}_{n}}\,{\cal A}
\,+
\biggl(\frac{\partial\hspace{0.04cm}\Omega}
{\partial\hspace{0.02cm}\bar{\xi}^{\,\prime}_{n}}\biggr)_{\!\bar{\xi}^{\,\prime}_{n}
\hspace{0.02cm}=\hspace{0.03cm}
\partial\hspace{0.01cm}{\cal A}/\partial\hspace{0.02cm}\xi_{n}}
\!\!\!+\, \xi_{n}\biggr\}\hspace{0.02cm}
\langle\hspace{0.02cm}\bar{\xi}^{\,\prime}\hspace{0.02cm}|\,\xi\hspace{0.02cm}\rangle.
\label{eq:8o}
\end{equation}
In the second term on the right-hand side instead of variables $\bar{\xi}_{n}^{\prime}$ in the derivative (\ref{eq:7d}) it is necessary to substitute $\partial\hspace{0.02cm}{\cal A}/\partial \hspace{0.02cm}\xi_n$.\\
\indent Finally, we consider the remaining term in (\ref{eq:4w}) containing the product $\hat{A}\hspace{0.02cm}[\hspace{0.03cm}a^{\phantom{+}}_{0}, a^{+}_{n}\hspace{0.02cm}]$. We present the matrix element of this product similar to (\ref{eq:7w}) in the following form:
\begin{equation}
\langle\hspace{0.02cm}\bar{\xi}^{\,\prime}\hspace{0.02cm}|\hspace{0.03cm}
\hat{A}\hspace{0.02cm}[\hspace{0.03cm}a^{\phantom{+}\!}_{0}, a^{+}_{n}\hspace{0.02cm}]|\,\xi\hspace{0.02cm}\rangle
=
\langle\hspace{0.02cm}\bar{\xi}^{\,\prime}\hspace{0.02cm}|\hspace{0.03cm}
[\hspace{0.03cm}a^{\phantom{+}}_{0}, a^{+}_{n}\hspace{0.02cm}]\hspace{0.02cm}\hat{A}|\,\xi\hspace{0.02cm}\rangle
+
\langle\hspace{0.02cm}\bar{\xi}^{\,\prime}\hspace{0.02cm}|\hspace{0.03cm}
[\hspace{0.02cm}\hat{A},[\hspace{0.03cm}a^{\phantom{+}}_{0}, a^{+}_{n}\hspace{0.02cm}]\hspace{0.02cm}]|\,\xi\hspace{0.02cm}\rangle.
\label{eq:8p}
\end{equation}
We perform analysis of the first term in the same way as it was just done for the matrix element 
$\langle\hspace{0.02cm}\bar{\xi}^{\,\prime}\hspace{0.02cm}|\hspace{0.03cm}
\hat{A}\hspace{0.02cm}[\hspace{0.03cm}a^{\phantom{+}\!}_{0}, a^{-}_{n}\hspace{0.02cm}]|\,\xi\hspace{0.02cm}\rangle$. 
The last step is to use the expressions (\ref{eq:7l}) and (\ref{eq:7x}). Collecting similar terms and recalling the definition of the derivative $\partial\hspace{0.04cm}\Omega/\partial\hspace{0.03cm} \xi_{n}$, Eq.\,(\ref{eq:7p}), we can write the expression above in the form similar to (\ref{eq:8o})
\begin{equation}
\langle\hspace{0.02cm}\bar{\xi}^{\,\prime}\hspace{0.02cm}|\hspace{0.03cm}
[\hspace{0.03cm}a^{\phantom{+}\!}_{0}, a^{+}_{n}\hspace{0.02cm}]\hspace{0.02cm}\hat{A}|\,\xi\hspace{0.02cm}\rangle
=
\biggl\{-\frac{\partial\hspace{0.04cm}\Omega}{\partial\hspace{0.02cm}\xi_{n}}\,{\cal A}
\,+
\biggl(\frac{\partial\hspace{0.04cm}\Omega}
{\partial\hspace{0.02cm}\xi_{n}}\biggr)_{\!\xi_{n}
\hspace{0.02cm}=\hspace{0.03cm}
\partial{\cal A}/\partial\hspace{0.02cm}\bar{\xi}^{\,\prime}_{n}}
\!\!\!-\, \bar{\xi}^{\,\prime}_{n}\biggr\}\hspace{0.02cm}
\langle\hspace{0.02cm}\bar{\xi}^{\,\prime}\hspace{0.02cm}|\,\xi\hspace{0.02cm}\rangle.
\label{eq:8a}
\end{equation}
Here, in the second term instead of the variables $\xi_n$ in the derivative (\ref{eq:7p}) it is necessary to substitute $\partial\hspace{0.03cm}{\cal A}/\partial\hspace{0.02cm} \bar{\xi}_{n}^{\prime}$.\\
\indent It only remains to analyze the last term in (\ref{eq:8p}). Taking into account (\ref{eq:8u}), we rewrite the double commutator as follows:
\begin{align}
[\hspace{0.02cm}\hat{A},[\hspace{0.03cm}a^{\phantom{+}\!}_{0}, a^{+}_{n}\hspace{0.02cm}]\hspace{0.02cm}]
&=
-\beta\hspace{0.02cm}[\hspace{0.02cm}a_{0},
[\hspace{0.02cm}a^{\phantom{+}\!}_{0},a^{+}_{n}\hspace{0.02cm}]\hspace{0.03cm}]
+
\gamma\hspace{0.02cm}[\hspace{0.02cm}(a_{0})^{2}, [\hspace{0.02cm}a^{\phantom{+}\!}_{0},a^{+}_{n}\hspace{0.02cm}]\hspace{0.03cm}]
=
-\beta\hspace{0.02cm}a^{+}_{n} + \gamma\hspace{0.02cm}\{\hspace{0.02cm}a^{\phantom{+}\!}_{0},a^{+}_{n}\hspace{0.02cm}\}
=
\label{eq:8s} \\[1ex]
&=
-\beta\hspace{0.02cm}a^{+}_{n} + 2\hspace{0.02cm}\gamma\hspace{0.03cm}a^{+}_{n}a^{\phantom{+}}_{0} + \gamma\hspace{0.03cm}[\hspace{0.03cm}a^{\phantom{+}\!}_{0},a^{+}_{n}\hspace{0.02cm}].
\notag
\end{align}
Here, we have used the commutation rule \eqref{ap:A4}. In view of (\ref{eq:7i}) and (\ref{eq:7a}), we get
\begin{equation}
\langle\hspace{0.02cm}\bar{\xi}^{\,\prime}\hspace{0.02cm}|\hspace{0.03cm}
[\hspace{0.02cm}\hat{A},[\hspace{0.03cm}a^{\phantom{+}\!}_{0}, a^{+}_{n}\hspace{0.02cm}]\hspace{0.02cm}]\hspace{0.02cm}|\,\xi\hspace{0.02cm}\rangle
=
\biggl\{\bar{\xi}^{\,\prime}_{n}\hspace{0.02cm}\bigl(-\beta + 2\hspace{0.02cm}\gamma\hspace{0.04cm}\Omega\bigr)
-
\gamma\hspace{0.02cm}\biggl(\frac{\partial\hspace{0.04cm}\Omega}
{\partial\hspace{0.02cm}\xi_{n}}\biggr)\!\hspace{0.01cm}\biggr\}\hspace{0.02cm}
\langle\hspace{0.02cm}\bar{\xi}^{\,\prime}\hspace{0.02cm}|\,\xi\hspace{0.02cm}\rangle.
\label{eq:8d}
\end{equation}
Now we can write out in full the expression for the matrix element (\ref{eq:4w}). Substituting the obtained matrix elements (\ref{eq:7q}), (\ref{eq:7z}), (\ref{eq:8o}), (\ref{eq:8p}) with (\ref{eq:8a}) and (\ref{eq:8d}) into (\ref{eq:4w}), we derive
\begin{equation}
\langle\hspace{0.02cm}(k)^{\prime}_{p}\hspace{0.02cm}|\hspace{0.03cm}
\hat{A}\hspace{0.03cm}\hat{\eta}_{\mu}(z)\hspace{0.02cm} \hat{D}_{\mu}\hspace{0.01cm}|\hspace{0.02cm}(k - 1)_{x}\hspace{0.02cm}\rangle
=
\label{eq:8f}
\end{equation}
\[
=
-\frac{i}{2}\,\biggl\{{\cal A}\hspace{0.02cm}
\sum\limits^{2}_{n\hspace{0.02cm}=\hspace{0.02cm}1}
\,\Bigl[\hspace{0.03cm}\Xi^{(\hspace{0.02cm}k-1,\hspace{0.03cm}k)}_{\bar{n}}(z)
\bigl(p^{(k)}_{\bar{n}} - e\hspace{0.02cm}A^{\phantom{(k)}\!\!\!\!\!}_{\bar{n}}(x^{(k-1)})\bigr)
+
\bar{\Xi}^{(\hspace{0.02cm}k,\hspace{0.03cm}k-1)}_{\hspace{0.02cm}n}(z)\hspace{0.04cm}
\bigl(p^{(k)}_{n} - e\hspace{0.02cm}A^{\phantom{(k)}\!\!\!\!\!}_{n}(x^{(k-1)})\bigr)\Bigr]
-
\]
\[
-\,\biggl(1 + \frac{1}{2}\,z\!\hspace{0.03cm}\biggr)
\frac{\!\!\!\partial{\cal A}}{\partial\hspace{0.02cm}\xi^{(k-1)}_{n}}\,
\bigl(p^{(k)}_{n} - e\hspace{0.02cm}A^{\phantom{(k)}\!\!\!\!\!}_{n}(x^{(k-1)})\bigr)
\,-
\]
\[
\begin{split}
-\,
z\hspace{0.03cm}\biggl(\frac{i\sqrt{3}}{2}\hspace{0.02cm}\biggr)\hspace{0.02cm}\biggl\{
&\biggl[\biggl(\frac{\partial\hspace{0.04cm}\Omega}
{\partial\hspace{0.02cm}\bar{\xi}^{\,\prime (k)}_{n}}\biggr)_{\!\bar{\xi}^{\,\prime (k)}_{n}
\hspace{0.02cm}=\hspace{0.03cm}
\partial\hspace{0.01cm}{\cal A}/\partial\hspace{0.02cm}\xi^{(k-1)}_{n}}
\!\!+\, \xi^{(k-1)}_{n}\biggr]
\bigl(p^{(k)}_{\bar{n}} - e\hspace{0.02cm}A^{\phantom{(k)}\!\!\!\!\!}_{\bar{n}}(x^{(k-1)})\bigr)
\,+ \\[1ex]
+
&\biggl[\biggl(\frac{\partial\hspace{0.04cm}\Omega}
{\partial\hspace{0.02cm}\xi^{(k-1)}_{n}}\biggr)_{\!\xi^{(k-1)}_{n}
\hspace{0.02cm}=\hspace{0.03cm}
\partial{\cal A}/\partial\hspace{0.02cm}\bar{\xi}^{\,\prime (k)}_{n}}
\!\!-\, \bar{\xi}^{\,\prime (k)}_{n}\biggr]
\bigl(p^{(k)}_{n} - e\hspace{0.02cm}A^{\phantom{(k)}\!\!\!\!\!}_{n}(x^{(k-1)})\bigr)\!\biggr\}
\,- \\[1ex]
-\,
z\hspace{0.03cm}\biggl(\frac{i\sqrt{3}}{2}\hspace{0.02cm}\biggr)\hspace{0.02cm}
&\biggl[\hspace{0.03cm}\bar{\xi}^{\,\prime}_{n}\hspace{0.02cm}\bigl(-\beta + 2\hspace{0.02cm}\gamma\hspace{0.04cm}\Omega\hspace{0.02cm}\bigr)
-
\gamma\hspace{0.02cm}\biggl(\frac{\partial\hspace{0.04cm}\Omega}
{\partial\hspace{0.02cm}\xi_{n}}\biggr)\!\hspace{0.03cm}\biggr]
\bigl(p^{(k)}_{n} - e\hspace{0.02cm}A^{\phantom{(k)}\!\!\!\!\!}_{n}(x^{(k-1)})\bigr)\!\biggr\}
\hspace{0.03cm}\times
\end{split}
\]
\[
\times
\hspace{0.03cm}
\langle\hspace{0.02cm}\bar{\xi}^{\,\prime(k)}\hspace{0.02cm}|\,\xi^{(k-1)}
\hspace{0.02cm}\rangle\hspace{0.02cm}
\langle\hspace{0.02cm}p^{(k)}|\,x^{(k-1)\hspace{0.02cm}}\rangle.
\]
Here, we have introduced the notations (cp. with (\ref{eq:2u}))
\[
\begin{split}
&\Xi^{(\hspace{0.02cm}k-1,\hspace{0.04cm}k)}_{\bar{n}}(z)
=
\biggl(1 + \frac{1}{2}\,z\!\hspace{0.03cm}\biggr)\hspace{0.02cm}\xi^{(k-1)}_{\bar{n}}
+
z\hspace{0.03cm}\biggl(\frac{i\sqrt{3}}{2}\hspace{0.02cm}\biggr)\!\hspace{0.02cm}
\biggl(\displaystyle\frac{\!\partial\hspace{0.04cm}\Omega}
{\partial\hspace{0.03cm}\bar{\xi}^{\hspace{0.03cm}\prime (k)}_{n}}\biggr),
\\[1ex]
&\overline{\Xi\!}^{\;(\hspace{0.02cm}k,\hspace{0.04cm}k-1)}_{\;n}(z)
=
\biggl(1 + \frac{1}{2}\,z\!\hspace{0.03cm}\biggr)\hspace{0.02cm}
\bar{\xi}^{\hspace{0.02cm}\prime(k)}_{n}
+
z\hspace{0.03cm}\biggl(\frac{i\sqrt{3}}{2}\hspace{0.02cm}\biggr)\!\hspace{0.02cm}
\biggl(\displaystyle\frac{\!\!\!\partial\hspace{0.03cm}\Omega}
{\partial\hspace{0.03cm}\xi^{(k-1)}_{n}}\biggr).
\end{split}
\]
The first term on the right-hand side of (\ref{eq:8f}) with the function ${\cal A} = {\cal A}(\bar{\xi}^{\hspace{0.02cm}\prime (k)}, \xi^{(k-1)})$ has a quite reasonable form. On the structure it corresponds to the initial operator expression $\hat{A}\hspace{0.02cm}\hat{\eta}_{\mu}(z)\hat{D}_{\mu}$ in (\ref{eq:3q}). The remaining terms are connected with the presence of additional commutators on the right-hand sides of (\ref{eq:7w}) and (\ref{eq:8p}), which inevitably violate symmetry of the expressions with respect to the creation $a_{n}^{+}$ and annihilation $a_{n}^{\!-}$ operators. The consequence of this is appearing the terms in (\ref{eq:8f}) of the type $(\partial\hspace{0.04cm} \Omega/\partial\hspace{0.03cm}\bar{\xi}_{n}^{\hspace{0.02cm}\prime})_{\bar{\xi}_{n}^{\,\prime}\hspace{0.02cm}=\hspace{0.03cm}\partial {\cal A}/\partial\hspace{0.02cm}\xi_n}$, which cannot be easily interpreted. In Part II we will consider somewhat different formalism which enables us at least on the formal level to write the expression (\ref{eq:8f}) in a more symmetric and visual form.


\section{Conclusion}
\setcounter{equation}{0}
\label{section_9}

In this paper we have taken initial steps to develop a formalism needed to construct the path integral representation for the Green's function of a massive vector particle within the framework of the Duffin-Kemmer-Petiau theory with deformation. One of the key point in our approach is the use of the connection between the deformed DKP-algebra and an extended system of parafermion trilinear commutation relations for the creation and annihilation operators $a_{k}^{\pm}$ obeying para-Fermi statistics of order 2 and for an additional operator $a_{0}$. We recall that the latter appears as some additional abstract element of the algebra  $\mathfrak{so}(2M+2)$. We have suggested an explicit representation of the operator $a_{0}$ constructed from the generators of the group $SO(2M)$. \\
\indent We have calculated all necessary matrix elements, which will be used in analysis of the contributions of the second and third orders with respect to the covariant derivative $\hat{D}_{\mu}$ in generalized Hamiltonian (\ref{eq:3p}). Although these matrix elements are presented in the most compact and visual form, the final expression for the whole matrix element of the contribution linear in the covariant derivative, Eq.\,(\ref{eq:8f}), ultimately proved to be cumbersome. One of the purposes of our next paper \cite{part_II} is to give to the obtained expression a more symmetric and simple form.



\begin{appendices}
\numberwithin{equation}{section}


\section{\bf Lie algebra \texorpdfstring{$\mathfrak{so}(2M+2)$}{a0a2}}
\numberwithin{equation}{section}
\label{appendix_A}

The Lie algebra of the orthogonal group $SO(2M + 2)$ has the following form:
$$
[\hspace{0.02cm}I_{\mu\nu},I_{\lambda\sigma}] =
\delta_{\nu\lambda}I_{\mu\sigma} + \delta_{\mu\sigma}I_{\nu\lambda}
-
\delta_{\mu\lambda}I_{\nu\sigma} - \delta_{\nu\sigma}I_{\mu\lambda}
$$
with $I_{\mu\nu} = -I_{\nu\mu}$. The indices $\mu,\nu,\ldots$ run values $1, 2,\hspace{0.02cm}\ldots\hspace{0.02cm},2M + 2$.
We introduce a new set of operators $\beta_{\mu}$ by setting\footnote{\hspace{0.02cm}We have redefined the operators $\beta_{\mu}$ from \cite{geyer_1968} for our case as follows:
\[
\beta_{\mu} \rightarrow 2\hspace{0.02cm}\beta_{\mu}\;\; \mbox{for}\;\; \mu = 1, 2,\hspace{0.02cm}\ldots\hspace{0.02cm}, 2M + 1.
\]
}
$$
\beta_{\mu} = -i\hspace{0.02cm}I_{\mu\, 2M+2}.
$$
Here the index $\mu$ runs values $1, 2,\hspace{0.02cm}\ldots\hspace{0.02cm},2M + 1$. The quantities $\beta_{\mu}$ are Hermitian
\begin{equation}
\beta^{\dagger}_{\mu} = \beta_{\mu}
\label{ap:A1}
\end{equation}
and obey the commutation relations
\[
\begin{split}
&[\hspace{0.02cm}\beta_{\mu},\beta_{\nu}] = I_{\mu\nu},\\[1ex]
&[\hspace{0.03cm}[\hspace{0.03cm}\beta_{\mu},\beta_{\nu}],\beta_{\lambda}\hspace{0.02cm}]
= \beta_{\mu}\hspace{0.02cm}\delta_{\nu\lambda} - \beta_{\nu}\hspace{0.02cm}\delta_{\mu\lambda}.
\end{split}
\]
The property (\ref{ap:A1}) enables us to introduce the Hermitian conjugate operators
\begin{equation}
\begin{split}
&a^{-}_{k} = \beta^{\phantom{-}\!\!}_{2\hspace{0.01cm}k - 1} - i\hspace{0.02cm}\beta^{\phantom{-}\!\!}_{2\hspace{0.01cm}k},\\[1ex]
&a^{+}_{k} = \beta^{\phantom{\dagger}\!}_{2\hspace{0.01cm}k - 1} + i\hspace{0.01cm}\beta^{\phantom{\dagger}\!}_{2\hspace{0.01cm}k},
\end{split}
\label{ap:A2}
\end{equation}
where $k = 1,2,\ldots,M$, and in addition to the $a^{\pm}_{k}$, a further operator is defined as 
\begin{equation}
a_{0} =\beta_{2M + 1}\, \Bigl(\equiv -2\hspace{0.02cm}i\hspace{0.03cm}I_{2M + 1\;2M + 2}\Bigr).
\label{ap:A3}
\end{equation}
The commutation relations between the operators $a^{\pm}_{k}$ are
\begin{align}
&[\hspace{0.02cm}a^{\pm}_{k}, [\hspace{0.02cm}a^{\mp}_{m},
a^{\pm}_{n}\hspace{0.02cm}]\hspace{0.02cm}] = 2\hspace{0.02cm}\delta^{\phantom{\dagger}\!}_{km}\hspace{0.02cm}a^{\pm}_{n},
\notag\\[1ex]
&[\hspace{0.02cm}a^{\pm}_{k}, [\hspace{0.02cm}a^{\pm}_{m},
a^{\pm}_{n}\hspace{0.02cm}]\hspace{0.02cm}] = 0,
\notag\\[1ex]
&[\hspace{0.02cm}a^{\pm}_{k}, [\hspace{0.02cm}a^{\mp}_{m},
a^{\mp}_{n}\hspace{0.02cm}]\hspace{0.02cm}] = 2\hspace{0.02cm}\delta^{\phantom{\dagger}\!}_{km}\hspace{0.02cm}a^{\mp}_{n}
-
2\hspace{0.02cm}\delta^{\phantom{\dagger}\!}_{kn}\hspace{0.02cm}a^{\mp}_{m},
\notag
\end{align}
the commutation relations involving one operator $a_{0}$ are
\begin{align}
&[\hspace{0.02cm}a^{\pm}_{k}, [\hspace{0.02cm}a^{\mp}_{m},
a^{\phantom{\pm}\!\!}_{0}\hspace{0.02cm}]\hspace{0.02cm}]
=
2\hspace{0.02cm}\delta^{\phantom{\dagger}\!}_{km}\hspace{0.02cm}a^{\phantom{\pm}\!\!}_{0},
\notag\\[1ex]
&[\hspace{0.02cm}a^{\pm}_{k}, [\hspace{0.02cm}a^{\pm}_{m},
a^{\phantom{\pm}\!\!}_{0}\hspace{0.02cm}]\hspace{0.02cm}] = 0,
\notag\\[1ex]
&[\hspace{0.02cm}a^{\phantom{\pm}\!\!}_{0}, [\hspace{0.02cm}a^{\pm}_{k},
a^{\mp}_{m}\hspace{0.02cm}]\hspace{0.02cm}] = 0,
\notag\\[1ex]
&[\hspace{0.02cm}a^{\phantom{\pm}\!\!}_{0}, [\hspace{0.02cm}a^{\pm}_{k},
a^{\pm}_{m}\hspace{0.02cm}]\hspace{0.02cm}] = 0,
\notag
\end{align}
and the commutation relation involving two operators $a_{0}$ is
\begin{equation}
[\hspace{0.02cm}a^{\phantom{\pm}\!\!}_{0}, [\hspace{0.02cm}a^{\phantom{\pm}\!\!}_{0},
a^{\pm}_{k}\hspace{0.02cm}]\hspace{0.02cm}] = 4\hspace{0.02cm}a^{\pm}_{k}.
\label{ap:A4}
\end{equation}
\indent Further, the uniqueness conditions of vacuum state $|\hspace{0.03cm}0\rangle$ in the parastatistics of order $p$ are \cite{green_1953}:
\[
a^{-}_{k}\hspace{0.02cm}|\hspace{0.03cm}0\rangle = 0, \quad
\mbox{for all}\; k
\]
and
\[
\hspace{0.1cm}
a^{-}_{k} a^{+}_{l}|\hspace{0.03cm}0\rangle = p\hspace{0.04cm} \delta^{\phantom{+}\!}_{kl} |\hspace{0.03cm}0\rangle,
\quad\;\;
\mbox{for all}\; k,\,l.
\]
The relation
\begin{equation}
a_{0} |\hspace{0.03cm}0\rangle = \pm\hspace{0.02cm}p\hspace{0.03cm}|\hspace{0.03cm}0\rangle
\label{ap:A5}
\end{equation}
will be a consequence of requiring the uniqueness of the vacuum state. Note that the sign on the right-hand side of \eqref{ap:A5} may be chosen arbitrarily. Action of the operator $a_{0}$ on an arbitrary state vector
$$
|\hspace{0.03cm}ij\,k\ldots rs\rangle  = a^{+}_{i}a^{+}_{j}a^{+}_{k}\ldots
a^{+}_{r}a^{+}_{s} |\hspace{0.04cm}0\rangle
$$
is defined by the following formula:
$$
a_{0}|\hspace{0.03cm}ij\hspace{0.03cm}k\ldots rs\rangle =
$$
$$
= \pm\hspace{0.03cm} p\,|\hspace{0.03cm}ij\hspace{0.03cm}k\ldots rs\rangle
\mp 2\Bigl(\hspace{0.02cm} |\hspace{0.03cm}j\hspace{0.03cm}k\ldots rsi\hspace{0.01cm}\,\rangle
+
|\hspace{0.03cm}i\hspace{0.03cm}k\ldots rsj\,\rangle
+
\,\ldots\, + |\hspace{0.03cm}ij\hspace{0.03cm}k\ldots s\,r\,\rangle + |\hspace{0.03cm}ij\hspace{0.04cm}k\ldots rs\,\rangle
\Bigr).
$$
In particular, this implies in addition to (\ref{ap:A5})
\begin{align}
&a_{0} |\hspace{0.03cm}r\rangle = \pm\hspace{0.03cm}
(p - 2)\hspace{0.03cm}|\hspace{0.03cm}r\rangle, \notag \\[1ex]
&a_{0} |\hspace{0.03cm}kr\rangle = \pm\hspace{0.03cm}
(p - 2)\hspace{0.03cm}|\hspace{0.03cm}kr\rangle \mp 2\hspace{0.03cm} |\hspace{0.03cm}rk\rangle,
\label{ap:A6} \\[1ex]
&a_{0} |\hspace{0.03cm}jkr\rangle = \pm\hspace{0.03cm}
(p - 2)\hspace{0.03cm}|\hspace{0.03cm}j\hspace{0.02cm}kr\rangle \mp 2\hspace{0.03cm} \bigl(|\hspace{0.03cm}j\hspace{0.02cm}rk\rangle
+ |\hspace{0.03cm}kr\!\hspace{0.02cm}j\rangle\hspace{0.02cm}\bigr),
\notag \\[1ex]
&a_{0} |\hspace{0.03cm}ijkr\rangle = \pm\hspace{0.03cm}
(p - 2)\hspace{0.03cm}|\hspace{0.03cm}ij\hspace{0.02cm}kr\rangle \mp 2\hspace{0.03cm}
\bigl(\hspace{0.02cm} |\hspace{0.03cm}ij\hspace{0.02cm}rk\rangle
+ |\hspace{0.03cm}ikr\!\hspace{0.02cm}j\rangle + |\hspace{0.03cm}j\hspace{0.02cm}kri\rangle\hspace{0.02cm}\bigr).
\notag
\end{align}
In the paper \cite{geyer_1968} a general relation for arbitrary values $p$ and $M$, which connects the operator $a_0$ with the operators $N_{1},\hspace{0.02cm}\ldots\hspace{0.02cm},N_{M}$ is given  (without a proof), where
$$
N^{\phantom{+}\!\!}_{k} =
\frac{1}{2}\,[\hspace{0.02cm}a^{+}_{k},a^{-}_{k}\hspace{0.02cm}].
$$
Let us write out the explicit form of the relations for the first two values $p$ in the case when $M = 2$:
\begin{align}
p = 1:\quad &a_{0} = 4\hspace{0.02cm}N_{1}\hspace{0.02cm}N_{2},
\notag \\[1.5ex]
p = 2:\quad &a^{2}_{0} =\! 2\hspace{0.03cm}
\Bigl\{ 1 + \bigl[\hspace{0.02cm}2\hspace{0.02cm}(N^{\phantom{2}\!\!}_{1})^{2} \!-\! 1\hspace{0.02cm}\bigr]
\bigl[\hspace{0.02cm}2\hspace{0.02cm}(N^{\phantom{2}\!\!}_{2})^{2} \!-\! 1\hspace{0.02cm}\bigr]\!\Bigr\}.
\label{ap:A7}  
\end{align}


\section{\bf Para-Grassmann numbers}
\label{appendix_B}
\numberwithin{equation}{section}

In this Appendix we will list the most important formulas of commutation and differentiation with para-Grassmann numbers. We follow the definition of a para-Grassmann algebra suggested by Omote and Kamefuchi \cite{omote_1979}, namely a set of independent numbers $\xi_{1}, \xi_{2}, \hspace{0.02cm}\ldots\hspace{0.02cm},\xi_{M}$ are said to form a para-Grassmann algebra of order $p$ when these numbers satisfy the following relations:
\begin{equation}
\begin{split}
&[\hspace{0.03cm}\xi_{i}\hspace{0.02cm}, [\hspace{0.02cm}\xi_{j}\hspace{0.02cm},\xi_{k}\hspace{0.02cm}]\hspace{0.02cm}]
= 0,\\[1ex]
&\{\xi_{i_{1}},\xi_{i_{2}},\,\ldots\,,\xi_{i_{m}}\} = 0\quad
\mbox{for} \; m\geq p+1,
\end{split}
\label{ap:B1}
\end{equation}
where $i's,j,k = 1,2,\,\ldots\,,M$ and by the symbol $\{\xi_{i_{1}},\xi_{i_{2}},\,\ldots\,,\xi_{i_{m}}\}$ one means a product of $m$ $\xi$-numbers completely symmetrized with respect to the indices $i_{1},i_{2},\,\ldots\,,i_{m}$. For the special case $p = 2$ these relations are reduced to
\begin{equation}
\xi_{i}\hspace{0.02cm}\xi_{j}\hspace{0.02cm}\xi_{k}\hspace{0.02cm}
+
\hspace{0.02cm}\xi_{k}\hspace{0.02cm}\xi_{j}\hspace{0.02cm}\xi_{i} = 0.
\label{ap:B2}
\end{equation}
\indent Further, let us write out the rules of commutation between the para-Grassmann numbers and the creation and annihilation para-Fermi operators $a^{\pm}_{i}$:
\begin{align}
&[\hspace{0.03cm}a^{\pm}_{i}, [\hspace{0.02cm}a^{\mp}_{j},\xi^{\phantom{\pm}\!}_{k}\hspace{0.02cm}]\hspace{0.015cm}]
=
2\hspace{0.03cm}\delta^{\phantom{\pm}\!}_{i\!\hspace{0.02cm}j}\,\xi^{\phantom{\pm}\!}_{k},
\label{ap:B3} \\[1ex]
&[\hspace{0.03cm}a^{\pm}_{i}, [\hspace{0.02cm}a^{\pm}_{j},\xi^{\phantom{\pm}\!}_{k}\hspace{0.02cm}]\hspace{0.015cm}]
= 0,
\label{ap:B4} \\[1ex]
&[\hspace{0.03cm}\xi_{i}, [\hspace{0.02cm}\xi_{j},a^{\pm}_{k}\hspace{0.02cm}]\hspace{0.02cm}] = 0.
\label{ap:B5}
\end{align}
These relations hold for parastatistics of an arbitrary order $p$.\\
\indent Let us present the formulas of differentiation with respect to a para-Grassmann number $\xi$. Throughout this text we mean left differentiation. The required formulas are \cite{ohnuki_1980}
\begin{align}
&\frac{\partial\hspace{0.02cm}\bigl(\hspace{0.02cm}[\hspace{0.03cm}\xi,\zeta\hspace{0.03cm}]
\hspace{0.04cm}
g(\xi)\!\hspace{0.04cm}\bigr)}{\partial\hspace{0.03cm}\xi}
=
\biggl(\frac{\partial\hspace{0.03cm}
[\hspace{0.03cm}\xi,\zeta\hspace{0.03cm}]}{\partial\hspace{0.03cm}\xi}\biggr)
\hspace{0.02cm}g(\xi)
+
[\hspace{0.03cm}\xi,\zeta\hspace{0.03cm}]\,
\frac{\partial\hspace{0.02cm}g(\xi)}{\partial\hspace{0.03cm}\xi}\,, 
\label{ap:B6} \\[1ex]
&\frac{\partial\hspace{0.03cm}}{\partial\hspace{0.03cm}\xi}\;
[\hspace{0.03cm}\xi,\zeta\hspace{0.03cm}]
= 2\hspace{0.03cm}\zeta. 
\label{ap:B7}
\end{align}

\end{appendices}

\end{document}